\documentclass[12pt]{article}
\usepackage{amssymb, amsmath, amsfonts,latexsym, verbatim, amsthm, mathrsfs}

\title{Quantum Mechanics in Multiply-Connected Spaces}
\author{
Detlef D\"urr\footnote{Mathematisches Institut der Universit\"{a}t
     M\"{u}nchen, Theresienstra{\ss}e 39, 80333 M\"{u}nchen, Germany.
     E-mail: duerr@mathematik.uni-muenchen.de},
Sheldon Goldstein\footnote{Departments of Mathematics, Physics and
     Philosophy, Hill Center, Rutgers, The State University of New  
     Jersey, 110 Frelinghuysen Road, Piscataway, NJ 08854-8019, USA.
     E-mail: oldstein@math.rutgers.edu},
James Taylor\footnote{Center for Talented Youth, Johns Hopkins
    University, McAuley Hall, Suite 400, 5801 Smith Ave, Baltimore,
    MD 21209, USA. E-mail: jostylr@member.ams.org},\\
Roderich Tumulka\footnote{Mathematisches Institut,
     Eberhard-Karls-Universit\"at, Auf der Morgenstelle 10, 72076
     T\"ubingen, Germany. E-mail:
     tumulka@everest.mathematik.uni-tuebingen.de},
\ and Nino Zangh\`\i\footnote{Dipartimento di Fisica dell'Universit\`a
     di Genova and INFN sezione di Genova, Via Dodecaneso 33, 16146
     Genova, Italy. E-mail: zanghi@ge.infn.it}
}
\date{June 27, 2006} 

\newcommand{\z}[1]{{#1}}

\newtheorem{assertion}{Assertion}

\newtheorem{defn}{Definition}
{\bf}{\it}

\newcommand{\dud}[2]{\ensuremath{\frac{\partial {#1}}{\partial {#2}}}}

\newcommand{\rvarn}[1]{\ensuremath{\mathbb{R}^{#1}}}

\newcommand{\sdud}[2]{\ensuremath{\frac{d{#1}}{d{#2}}}}

\newcommand{\gp}{\ensuremath{\psi}}
\newcommand{\gD}{\ensuremath{\Delta}}
\newcommand{\gr}{\ensuremath{\pi}}
\newcommand{\re}{\ensuremath{\mathbb{R}}}
\newcommand{\se}{Schr\"odinger's equation}
\newcommand{\sch}{Schr\"odinger}
\newcommand{\qpotc}[1]{\ensuremath{\frac{\hbar^2}{2m_{#1}}}}

\newcommand{\qmu}[1]{\ensuremath{\frac{\hbar}{m_{#1}}}}

\newcommand{\defi}{\ensuremath{:=}}

\newcommand{\seq}[1]{\ensuremath{i \hbar \dud{\gp}{t} = - \sum_{k =
1}^{#1} \qpotc{k} \gD_{k} \gp + V \gp   }}

\newcommand{\im}{\ensuremath{\mathrm{Im}}}
\newcommand{\inpr}[2]{\ensuremath{({#1}, {#2})}}
\newcommand{\dd}{\ensuremath{d}}
\newcommand{\rd}{\rvarn{\dd}}
\newcommand{\rdn}{\rvarn{\dd N}}
\newcommand{\nrd}{\ensuremath{{}^N\mspace{-1.0mu}\rvarn{\dd}}}

\newcommand{\defnrd}{\ensuremath{\{S|S \subseteq \rd, |S| = N \}}}

\newcommand{\wf}{wave function}

\newcommand{\covspa}{\ensuremath{\widehat{\gencon}}}
\newcommand{\proj}{\ensuremath{\gr   }}
\newcommand{\gencon}{\ensuremath{\mathcal{Q}}}
\newcommand{\fund}[1]{\ensuremath{\pi_1 ({#1})}}
\newcommand{\deckg}{\ensuremath{Cov(\covspa, \gencon)}}
\newcommand{\deckt}{\ensuremath{\sigma}}

\newcommand{\concov}{\ensuremath{\gamma_{\deckt}}}

\newcommand{\conc}[1]{\ensuremath{\gamma_{#1}}}
\newcommand{\cmplx}{\ensuremath{\mathbb{C}}}
\newcommand{\deck}[1]{\ensuremath{\deckt_{#1}}}

\newcommand{\baseq}{\ensuremath{q}}
\newcommand{\covq}{\ensuremath{\hat{q}}}

\newcommand{\canq}{\ensuremath{\boldsymbol{q}}}

\newcommand{\spins}{\ensuremath{\cmplx^{2s +1}}}
\newcommand{\sspa}{\ensuremath{W}}

\newcommand{\hb}{\ensuremath{\hbar}}
\newcommand{\bm}{Bohmian mechanics}

\newcommand{\Sect}[1]{Section {#1}}

\newcommand{\wh}[1]{\ensuremath{\widehat{#1}}}

\newcommand{\fiber}{covering fiber}

\newcommand{\eqv}{\ensuremath{\sim}}

\newcommand{\limts}{}
\newcommand{\concav}{\ensuremath{\gamma}}

\newcommand{\sidecom}[1]{}
\newcommand{\wdefi}{\ensuremath{=:}}
\newcommand{\asp}{Character Quantization Principle}
\newcommand{\herm}{Hermitian}

\newcommand{\id}{\ensuremath{\mathrm{Id}}}

\newcommand{\Hilbert}{\mathscr{H}}

\newcommand{\class}{\mathscr{C}} 
\newcommand{\RRR}{\mathbb{R}} 
\newcommand{\CCC}{\mathbb{C}} 
\newcommand{\ZZZ}{\mathbb{Z}} 
\newcommand{\Q}{\gencon} 
\newcommand{\Euclideanspace}{\mathscr{E}} 
\newcommand{\electric}{\boldsymbol{E}} 
\newcommand{\magnetic}{\boldsymbol{B}} 
\newcommand{\cylinder}{\mathcal{C}} 
\newcommand{\vA}{\boldsymbol{A}} 
\newcommand{\set}{\mathscr{S}} 
\newcommand{\rawH}{-\tfrac{\hbar^2}{2} \gD + V} %
\newcommand{\nodes}{\mathcal{N}} 
\newcommand{\holonomy}{\ensuremath{h}} 
\newcommand{\closedcurve}{\alpha}
\newcommand{\opencurve}{\beta}

\newcommand{\nrtre}{{}^N \RRR^3}
\newcommand{\covr}{\hat{r}}
\newcounter{remi} 

\begin{document}
\maketitle

\begin{abstract}
  We \z{explain} why, in a configuration space that is multiply connected,
  i.e., whose fundamental group is nontrivial, there are \emph{several}
  quantum theories, corresponding to different choices of topological
  factors.  We do this in the context of Bohmian mechanics, a quantum
  theory without observers from which the quantum formalism can be derived.
  What we do can be regarded as generalizing the Bohmian dynamics \z{on}
  $\RRR^{3N}$ to arbitrary Riemannian manifolds, and classifying the
  possible dynamics \z{that arise}.  This approach provides a new
  understanding of the topological features of quantum theory, such as the
  symmetrization postulate for identical particles.  For our analysis we
  employ wave functions on the universal covering space of the
  configuration space.

\medskip


\end{abstract}

\tableofcontents

\section{Introduction} \label{sec:intro}

We \z{shall be concerned here with} topological effects in quantum
mechanics, and shall elaborate on some results first described in
\cite{topid0}. The kind of statement on which we \z{shall} focus asserts
that if the configuration space $\gencon$ is a
multiply-connected\footnote{Recall that a manifold $\gencon$ is
\emph{simply connected} if all closed curves in $\gencon$ are contractible.
Otherwise, it is \emph{multiply connected}. (Note that ``multiply
connected'' is different from the notion ``$n$-connected'' for $n \geq 0$,
which is sometimes used in the literature on algebraic topology
\cite[p.~51]{Spa66} and means that the first $n$ homotopy groups
$\pi_n(\Q)$ are all trivial.) Examples of simply-connected spaces are
$\RRR^d$ for any $d \geq 0$, the spheres $S^d$ for $d \geq 2$, or the
punctured spaces $\RRR^d \setminus \{0\}$ for $d \geq 3$; examples of
multiply-connected spaces \z{are} the circle $S^1$, the torus $S^1 \times
S^1$, the punctured plane $\RRR^2 \setminus \{0\}$, or generally $\RRR^d
\setminus U$ where $U$ is a subspace of dimension $d-2$, $d \geq 2$.}
Riemannian manifold then there exist \emph{several} quantum theories in
$\gencon$. More precisely, the dynamics is not completely determined by
specifying $\gencon$ \z{(whose metric we regard as incorporating the
``masses of the particles'')} together with \z{the potential and the value
space of the wave function}; in addition, one can choose \emph{topological
factors}, which form a representation (or twisted representation) of the
fundamental group $\fund{\Q}$ of $\gencon$.  In each of the theories, the
Hamiltonian is locally equivalent to $\rawH$, though not globally. The
investigation in this paper is continued with other methods in three
follow-up papers \cite{topid1B, topid1C, topid1D}.

Our interest lies in explaining why there is more than one quantum
theory and how the several possibilities arise, and in classifying the
possibilities. The formulation of quantum mechanics we use for this
purpose is Bohmian mechanics \cite{Bohm52, Bell66, DGZ92, survey,
  DGZ96, Gol01}, a quantum theory without observers; it describes a
world in which particles have trajectories, guided by a wave function
$\psi_t$; observers in this world would find that the results of their
experiments obey the quantum formalism \cite{Bohm52, Bell66, DGZ92,
  DGZ03}. We will give a brief review of Bohmian mechanics in
\Sect{\ref{sec:bm}}.  Most of our mathematical considerations and
methods are equally valid, relevant, and useful in orthodox quantum
mechanics, or any other \z{version}  of quantum mechanics.  Bohmian mechanics,
however, provides a sharp mathematical justification of these
considerations that is \z{absent}  in the orthodox framework.

The motion of the configuration in a Bohmian $N$-particle system can
be regarded as \z{corresponding to}  a dynamical system in the configuration space $\Q =
\RRR^{3N}$, defined by a time-dependent vector field $v^{\psi_t}$ on
$\Q$ which in turn is defined, by the Bohmian law of motion, in terms
of $\psi_t$. We are concerned here with the analogues of the Bohmian
law of motion for the case that $\Q$ is, instead of $\RRR^{3N}$, an
arbitrary Riemannian manifold.\footnote{Manifolds will throughout be
  assumed to be Hausdorff, paracompact, connected, and $C^\infty$.
  They need not be orientable.}  The main result is that, if $\Q$ is
multiply connected, there are several such analogues: several Bohmian
dynamics, which we will describe in detail, corresponding to different
choices of the topological factors.

To define a Bohmian theory in a manifold $\Q$ ultimately amounts to
defining trajectories in $\Q$ and their probabilities. This leads to clear
mathematical classification questions, while from the orthodox point of
view the ground rules with respect to the issue of the existence of several
quantum theories with different topological factors are less clear.  We
will review the differences between the two viewpoints in
\Sect{\ref{sec:orthodox}}.

The topological factors consists of, in the simplest case, phase factors
associated with non-contractible loops in $\Q$, forming a
character\footnote{By a \emph{character} of a group we refer to what is
sometimes called a unitary multiplicative character, i.e., a one-dimensional
unitary representation of the group.} of the fundamental group
$\fund{\Q}$. All characters can be physically relevant; we emphasize this
because it is easy to overlook the multitude of dynamics by focussing too
much on just one, the simplest one, which we will define in
\Sect{\ref{sec:immediate}}: the \emph{immediate generalization} of the
Bohmian dynamics from $\RRR^{3N}$ to a Riemannian manifold, or, as we shall
briefly call it, the \emph{immediate Bohmian dynamics}.

Apart from the mathematical exercise, what do we gain from
studying the possible Bohmian dynamics on manifolds? 
\begin{itemize}
\item A new understanding of how topological factors in quantum
  mechanics can be \z{regarded}  as arising.
\item A presumably complete \emph{classification} of the topological
  factors in quantum mechanics, including some, corresponding to what
  we call \emph{twisted representations} of $\fund{\Q}$, that have
  not, to our knowledge, been considered \z{ so far}  in the literature.
\item An explanation of the fact that the wave function of a system of
identical particles is either symmetric or anti-symmetric, a fact that
\z{(at least insofar as nonrelativistic quantum mechanics is concerned)} is
usually, instead of being derived, introduced as a \emph{symmetrization
postulate}.  This application is discussed in detail in a sister paper to
this one \cite{topid2}, and will only be touched upon briefly here.
\end{itemize}
\z{Our main} motivation \z{for studying} the question of Bohmian dynamics
on manifolds was in fact the investigation of the symmetrization postulate
for identical particles.

As we have already mentioned, one of the different Bohmian dynamics on a
manifold $\Q$ is special, as it is the immediate Bohmian dynamics on
$\Q$.\footnote{That is, if one considers the value space of the wave
function as given. In \cite{topid1D}, in contrast, we obtain
several bundles of spin spaces from the Riemannian geometry of the
configuration space. We could take
   any of these bundles as the starting point for defining the
   dynamics, and then which one of the dynamics is immediate will
   depend on this choice.}
The other kinds of
Bohmian dynamics come in a hierarchy of increasing complexity. There are
three natural classes $\class_1, \class_2, \class_3$ of Bohmian dynamics,
related according to
\begin{equation}
  \class_0 \subseteq \class_1 \subseteq \class_2 \subseteq \class_3\,,
\end{equation}
where $\class_0$ contains only the immediate Bohmian dynamics.  The
dynamics of class $\class_1$, defined in \Sect{\ref{sec:scalarperiodic}},
involve topological phase factors forming a character of the fundamental
group $\fund{\Q}$. Those of class $\class_2$, defined in
\Sect{\ref{s:vectorvalued}} and in a more general setting in
\Sect{\ref{s:bundlevalued}}, involve topological factors that are \z{given
by} matrices, forming a \z{unitary representation of $\fund{\Q}$} or, in
the case of a vector bundle, a twisted representation (see the end of
\Sect{\ref{s:bundlevalued}} for the definition).  Those of class $\class_3$
will not be discussed here but in \cite{topid1B}; they involve changes in
connections and potentials that are not based on multiple connectivity.  As
we shall explain, the dynamics of bosons belongs to $\class_0$ while that
of fermions belongs to $\class_1$. More precisely, fermions can be regarded
as belonging either to $\class_0$, for a certain nontrivial vector bundle
defined in \cite{topid2} (for which bosons are of class $\class_1$), or to
$\class_1$, for the trivial bundle $\Q \times \CCC$. In \Sect{\ref{s:abs}}
we derive a dynamics of class $\class_1$ for the Aharonov--Bohm effect.  We
will define dynamics here in a non-rigorous way; a rigorous definition of
the classes $\class_0, \class_1, \class_2, \class_3$ is given in
\cite{topid1B}.

It is not obvious what ``all possible kinds of Bohmian dynamics''
should mean. We will investigate one approach here, while others, \z{as
already mentioned,}  are
studied in \cite{topid1B,topid1C,topid1D}. The present approach is based on
considering wave functions $\psi$ that are defined not on the
configuration space $\Q$ but on its universal covering space
$\covspa$. We then study which kinds of periodicity conditions,
relating the values on different levels of the covering fiber by a
topological factor, will ensure that the Bohmian velocity vector field
associated with $\psi$ is projectable from $\covspa$ to $\Q$.  This is
carried out in \Sect{\ref{sec:covering}} for scalar wave functions and
in \Sect{\ref{s:periodic2}} for wave functions with values in a
complex vector space (such as \z{a} spin-space) or a complex vector bundle.
In the case of vector bundles, we derive a novel kind of topological
\z{factor}, given by \z{a} twisted \z{representation}  of $\fund{\Q}$.

Let us mention the other approaches to defining the dynamics of classes
$\class_1, \class_2$, \z{and} $\class_3$. Since wave functions can be
regarded as sections of \herm\ bundles, i.e., complex vector bundles with a
connection and parallel Hermitian inner products, one approach
\cite{topid1B} considers all \herm\ bundles that are locally (but not
globally) isomorphic to a given one. Another approach \cite{topid1B}
expresses the dynamics in terms of the Hamiltonian and considers all
Hamiltonians $H$ that are locally (though \z{not necessarily} globally)
equivalent to $\rawH$. Another approach \cite{topid1D} regards the value space
of the wave function as a representation space (such as \z{a} spin-space)
of a suitable group (such as the rotation group) and classifies the \herm\
bundles consisting of representation spaces. A last approach \cite{topid1C}
removes a surface $\kappa$ from configuration space $\Q$, such that $\Q
\setminus \kappa$ is simply connected, and imposes a periodic boundary
condition relating the wave function on both sides of the new, ``virtual'',
boundary $\kappa$ by a topological factor.

In some cases, some of the classes coincide: When $\Q$ is simply connected,
then $\class_0 = \class_1 = \class_2$. When the wave function is a scalar
(as for spinless particles), then $\class_1 = \class_2 = \class_3$.  For
generic potentials, $\class_1 = \class_2 = \class_3$.

We encounter examples of multiply-connected configuration spaces in
two ways: either, as in the Aharonov--Bohm effect, by ignoring an
existing part of physical (or configuration) space, or, as for
identical particles or multiply-connected cosmologies, from the very
nature of the configuration space. In the former case the topological
factors of the \emph{effective} dynamics on the available
configuration space depend on external fields, while in the latter
case the topological factors of the \emph{fundamental} dynamics should
be compatible with any choice of external fields. This compatibility
is a strong restriction, which allows only the dynamics of class
$\class_1$.  Hence, as we shall argue more fully later, the several
fundamental quantum theories in $\Q$ are those given by $\class_1$.
This conclusion we call the \emph{\asp}, since the dynamics of
$\class_1$ are defined using the characters of the fundamental group
of $\Q$. It is formulated and discussed in \Sect{\ref{sec:asp}}.  We
conclude in \Sect{\ref{sec:conclusions}}.

The notion that multiply-connected spaces give rise to different quantum
theories is not new. Here is a sampling of the literature. A covering space
was used at least as early as 1950 by Bopp and Haag \cite{BH50} for the
configuration space of the spinning top; it was more fully \z{exploited} by
Dowker \cite{D72}, and \z{was} used by Leinaas and Myrheim \cite{LM77} for
the configuration space of identical particles.  Vector potentials on
multiply connected spaces were used by Aharonov and Bohm in
\cite{AB59}. Path integrals on multiply connected spaces began largely with
the work of Schulman \cite{S68,S71} and that of Laidlaw and DeWitt in
\cite{DL71}; see \cite{Sch81} for details. There is also the current
algebra approach of Goldin, Menikoff, and Sharp \cite{GMS81}.  Most of
these works are dedicated to scalar \wf s. The study of arbitrary manifolds
began with Laidlaw and DeWitt \cite{DL71}, which deals with path
integration on the universal covering space for scalar wave
functions. Nelson \cite{Nel85} derives the topological phase factors for
scalar wave functions from stochastic mechanics. Gamboa and Rivelles
\cite{GR91} consider relativistic Hamiltonians using a path-integral
approach.
Ho and Morgan \cite{HM96} provide a study of quantum
mechanics on $\RRR^d \times S^1$ for scalar wave functions.

\section{Perspective on Orthodox Quantum Mechanics}
\label{sec:orthodox}

When one considers a Bohmian dynamics and removes the Bohmian trajectories,
there still remains the wave function $\psi$, and a number of nontrivial
things can be said about it, such as, which space $\psi$ can be taken from
and how its evolution is defined.  As a consequence, much of the
mathematical discussion in this paper would be equally valid, applicable
and relevant for any other formulation of quantum mechanics. However, our
analysis of the emergence of the further kinds of wave functions, whose
main role in Bohmian mechanics is to define trajectories, would not work in
the same way if one were to dispense with the trajectories.

One would meet, when trying to carry out the program of this paper in
orthodox quantum mechanics, some difficulties that are absent in the
Bohmian approach. This is mainly because of two traits of Bohmian
mechanics: first, it is clear in the Bohmian framework  at which point the
specification of a theory is complete; and second, it is clear whether two
variants of a theory are physically equivalent or not.  Let us explain.

In the Bohmian framework, once the possible trajectories of all particles
have been defined (together with the appropriate equivariant probability
distribution, see Section \ref{sec:bm}) then the theory has been completely
specified, and there is neither need nor room for further axioms.  In
orthodox quantum mechanics, in contrast, it is not obvious what it is that
needs to be specified in order to have a variant of the theory.  The
Hilbert space $\Hilbert$ and the Hamiltonian $H$?  While they certainly
must be specified, they are certainly not enough.

One could think, for example, of different possible position observables in
the same Hilbert space, and these would lead to different predictions for
position measurements. Thus, one should specify, it would seem, $\Hilbert$,
$H$, and the operator, or commuting set of operators, for the position
observable. But would that be enough? Need we not also be told what
operator represents the momentum observable?  Need we not be told what
operators represent \emph{all} the observables?  This should be contrasted
with the fact that in Bohmian mechanics, once the dynamics of the particles
is specified, also the outcomes of all experiments are
specified.\footnote{One could argue that for exactly the same reason, to
specify the position observable in orthodox quantum mechanics \z{would be}
sufficient, as it \z{would fix} the statistics of \z{the} outcomes of every
measurement. This is true, and we think that this is a healthy
attitude. However, it is also quite against the spirit of orthodox quantum
mechanics which sets a high value on the ``democracy'' for all
observables.}

And what are, by the way, ``all'' observables? It seems clear that the list
of all observables should begin with position, momentum, and energy, but
where it should end is rather obscure.  In addition, the notion of
observable becomes somewhat problematic when the configuration space $\Q$
is a manifold. The problem is not so much that the position observable can
no longer be represented by a set of commuting position operators, as the
manifold may not permit global coordinates (e.g., on the circle); one
should conclude that the appropriate notion of position observable is then
a PVM (projection-valued measure) on $\Q$ acting on $\Hilbert$, associating
with every subset of $\Q$ a projection in $\Hilbert$. The more serious
problem concerns the momentum observable: already on the half-line, the
operator $p = -i\hbar \, d/dq$ does not have a self-adjoint extension.
More generally, on a Riemannian manifold $\Q$ the notion of the momentum
observable becomes \z{obscure,} as it is based on a translation symmetry
that may not exist in $\Q$. Thus, a momentum observable may not exist.
This brings us back to the point that it is not clear which observables
need to be specified in order to specify an orthodox quantum theory.

The contrast between the clarity of Bohmian mechanics and the vagueness of
orthodox quantum theory is perhaps even more striking when we consider the
issue of \z{the} physical equivalence of theories. In
this paper we shall always treat Bohmian theories, when they are
mathematically different but lead to the same trajectories (and
probabilities), as physically equivalent. For example, the dynamics we
shall define using wave functions on the covering space with the trivial
character is physically equivalent to the immediate Bohmian dynamics, using
wave functions on the configuration space.


In orthodox quantum mechanics, when should we regard two variants of the
theory as physically equivalent? The answer in the spirit of orthodox
quantum mechanics is, when they predict the same statistics for outcomes
for all experiments; that is, when they are empirically
equivalent.\footnote{This answer can be criticized on the grounds that
there are known examples of theories that are empirically equivalent though
physically inequivalent, such as Bohmian mechanics and stochastic mechanics
\cite{Nel85}, or the variants of Bohmian mechanics in which some of the
particles do not possess actual positions while their coordinates get
integrated over in the law of motion \cite{GTTZ04}.} This answer leads
again to the question, what are ``all'' observables? In addition, it leads
us to the possibly separate problem of identifying the observables of one
theory with the observables of another. Within the Bohmian framework, based
on a clear ontology and a correspondingly sharp specification of the
relevant physical structures and their behavior, no such questions and
problems can arise.

\section{Perspective on Spontaneous Collapse Theories}
\label{sec:GRW}
\newcommand{\vx}{\boldsymbol{x}}
\newcommand{\vq}{\boldsymbol{q}}
\renewcommand{\dd}{\ensuremath{3}}

Another approach besides Bohmian mechanics leading to quantum theories
without observers is that of spontaneous wave function collapse
\cite{Pearle76,GRW86,Bell87a, AGTZ06}; the simplest and best known model
of this kind is due to Ghirardi, Rimini, and Weber (GRW) \cite{GRW86}. Its
situation with respect to topological factors is very different from that
of Bohmian mechanics. For example, the situation of identical particles in
the GRW theory is different from that in Bohmian mechanics because the
latter is (in a suitable sense) automatically compatible with bosons and
fermions, whereas the equations of the GRW model require modification for
identical particles as follows \cite{DS95,Tum06}.

In the original GRW model (corresponding to $N$ distinguishable
particles), collapses are associated with points in 3-space and labels $i
\in \{1,\ldots,N\}$. Given the wave function $\psi: \RRR^{3N} \to \CCC$, a
collapse with label $i$ and location $\vx \in \RRR^3$ occurs with rate
\begin{equation}
   r_i(\vx|\psi) =  \langle \psi | \Lambda_i(\vx) \, \psi \rangle \,,
\end{equation}
where the collapse rate operator $\Lambda_i(\vx)$ is a multiplication
operator defined by
\begin{equation}
   \Lambda_i(\vx) \, \psi(\vq_1, \ldots, \vq_N) = \lambda \,
   \exp\Bigl(-\frac{(\vx-\vq_i)^2}{2a^2}\Bigr)
   \, \psi(\vq_1, \ldots, \vq_N)\,.
\end{equation}
The constants $\lambda$ and $a$ are parameters of the model. A collapse at
time $t$ and location $\vx$ with label $i$ changes the wave function
according to
\begin{equation}
   \psi_{t-} \mapsto \psi_{t+} = \frac{\Lambda_i(\vx)^{1/2} \psi_{t-}}
   {\|\Lambda_i(\vx)^{1/2} \psi_{t-}\|}\,.
\end{equation}

In the version for identical particles, collapses are associated with
locations $\vx$ only, without labels. Letting $\psi$ be either a symmetric or
an anti-symmetric function on $\RRR^{3N}$, a collapse occurs at location
$\vx \in \RRR^3$ with rate
\begin{equation}\label{GRWrate}
   r(\vx|\psi) =  \langle \psi | \Lambda(\vx) \, \psi \rangle \,,
\end{equation}
where
\begin{equation}
   \Lambda(\vx) = \sum_{i=1}^N \Lambda_i(\vx)\,,
\end{equation}
and changes $\psi$ according to
\begin{equation}\label{GRWcollapse}
   \psi \mapsto \frac{\Lambda(\vx)^{1/2} \psi}{\|\Lambda(\vx)^{1/2}
\psi\|}\,.
\end{equation}
Thus, the collapsed wave function is again symmetric respectively
anti-symmetric.

The arguments used in the present paper for deriving the topological
factors cannot be repeated in the context of the GRW theory because they
rely on particle configurations, which do not exist in the GRW theory. As a
consequence, indeed, configuration space does not play, in the GRW theory,
the same central role as in Bohmian mechanics, but merely that of a
convenient tool for representing the state vector as a function (similar to
the role, in Bohmian mechanics, of momentum space or of the set of spin
eigenvalues). Thus, it is hard to see how the GRW theory could provide any
reasons for the existence of several possibilities, corresponding to
different topological factors, in situations in which the configuration
space of the Bohmian theory is multiply connected. Moreover, for the GRW
theory, for which there are no particles to begin with, it is hard to see
why the multiply-connected natural configuration space $\nrd$ for $N$
identical particles, see Section \ref{sec:immediate}, should be considered
at all. Nonetheless, topological factors can always be introduced into GRW
theories, as in the example above, as we shall explain later in Section
\ref{is:rem}, Remark \ref{rem:GRW}.

\section{Bohmian Mechanics in $\rvarn{3N}$}
\label{sec:bm}

Bohmian mechanics is a theory about particles with definite locations.
The \z{theory specifies }  the trajectories in physical
space of these particles.  The object which determines the
trajectories is the \wf, familiar from quantum mechanics. More
precisely, the state of the system in \bm\ is given by the pair $(Q,
\gp)$; $Q = (\boldsymbol Q_1, \ldots, \boldsymbol Q_N) \in \rdn$ is
the configuration of the $N$ particles in our system and $\psi$ is a
(standard quantum mechanical) \wf\ on the configuration space \rdn,
taking values in some \emph{Hermitian vector space} $\sspa$, i.e., a
finite-dimensional complex vector space endowed with a
positive-definite Hermitian (i.e., conjugate-symmetric and
sesqui-linear) inner product $(\,\cdot\,,\,\cdot\,)$.  The state of
the system changes according to the guiding equation and \se:
\begin{equation} \label{e:bohm}
  \sdud{\boldsymbol{Q}_{k}}{t} = \qmu{k} \im \frac
  {\inpr{\gp}{\nabla_{k} \gp}}{ \inpr{\gp}{\gp}}
  (\boldsymbol{Q}_1,\ldots, \boldsymbol{Q}_N) \wdefi v_k^{\gp} (Q),
  \quad k= 1,\ldots,N
\end{equation}
\begin{equation}\label{e:sch01}
  \seq{N}
\end{equation}
where $V$ is the potential function with values \z{given by}  Hermitian
matrices (endomorphisms of $\sspa$). We call $\inpr{\phi(q)} {\psi(q)}$,
the inner product on the value space $\sspa$, the \emph{local inner
  product}, in distinction from the inner product $\langle \phi, \psi
\rangle$ on the Hilbert space of \wf s. For complex-valued \wf s, the
potential is a real-valued function on configuration space and the
local inner product is $\overline{\phi(q)}{\gp(q)}$, where the bar
denotes complex conjugation.

The empirical agreement between \bm\ and standard quantum mechanics is
grounded in equivariance \cite{DGZ92,DGZ03}.  In \bm, if the configuration
is initially random and distributed according to $|\gp_0|^2$, then the
evolution is such that the configuration at time $t$ will be
distributed according to $|\gp_t|^2$.  This property is called \z{the} 
equivariance of the $|\gp|^2$ distribution.  It follows from comparing
the transport equation
\begin{equation}\label{continuity}
  \dud{\rho_t}{t} =- \nabla \cdot (\rho_t v^{\gp_t}) 
\end{equation}
for the distribution $\rho_t$ of the configuration $Q_t$, where
$v^\psi = (v_1^\psi, \ldots, v_N^\psi)$, to \z{the quantum continuity
equation} 
\begin{equation}\label{dpsi2dt}
  \dud{|\gp_t|^2}{t} =- \nabla \cdot (|\gp_t|^2 v^{\gp_t}),
\end{equation}
which is a consequence of \sch's equation \eqref{e:sch01}. A
rigorous proof of equivariance requires showing that almost all (with
respect to the $|\gp|^2$ distribution) solutions of \eqref{e:bohm}
exist for all times. This was done in \cite{BDGPZ95,TT04}.  A more
comprehensive introduction to \bm\ may be found in \cite{Gol01,
survey, DGZ96}.

Spin is already incorporated in \eqref{e:bohm} and \eqref{e:sch01} if
one chooses for $\sspa$ \z{a}  suitable spin space \cite{Bell66}.  By
assumption, for one particle moving in \rvarn{3}, \sspa\ is a complex,
irreducible representation space of $SU(2)$, the universal covering
group\footnote{The universal covering space of a Lie group is again a
  Lie group, the \emph{universal covering group}. It should be
  distinguished from another group also called the \emph{covering
    group}: the group $Cov(\covspa,\gencon)$ of the covering (or deck)
  transformations of the universal covering space $\covspa$ of a
  manifold $\Q$, which will play an important role later.} of the
rotation group $SO(3)$.  If it is the spin-$s$ representation then
$\sspa = \spins$.

\section{The Immediate Generalization to Riemannian Manifolds}
\label{sec:immediate}

We now consider, in the role of the configuration space, a Riemannian
manifold $\Q$ instead of $\RRR^{3N}$. The primary physical motivation is
the study of identical particles, for which the natural configuration space
is the \z{set $\nrd$ of} all $N$-element subsets of $\RRR^3$,
\begin{equation}\label{nrddef}
  \nrd := \defnrd\,,
\end{equation}
which naturally carries the structure of a Riemannian manifold, in
fact a multiply-connected one. This configuration space was first
suggested in \cite{DL71} and \cite{LM77}; for further discussion see
\cite{topid2}.

But the generalization to manifolds is \z{also very natural
mathematically}. In addition, there are further cases of
physical relevance: One could consider, instead of $\RRR^3$, a curved
physical space. And in cases like the Aharonov--Bohm effect, the phase
shift that occurs can be attributed to the topology of the effectively
available configuration space, a subset of the entire configuration
space that can be viewed as a \z{multiply-connected} manifold.

\subsection{Euclidean Vector Spaces}

It is in fact easy to find a generalization of the Bohmian dynamics
to a Riemannian manifold $\Q$, which we call the immediate Bohmian
dynamics on $\Q$. One reason why it is so easy is that the law of
motion for the point $Q_t = Q(t) = (\mathbf{Q}_1(t), \ldots,
\mathbf{Q}_N(t))$ in \z{the} configuration space $\RRR^{3N}$ representing the
positions of all particles at time $t$ is almost independent of the
way in which $\RRR^{3N}$ is composed of $N$ copies of $\RRR^3$. In
fact, \eqref{e:bohm} can be written as
\begin{equation}\label{Qlawm}
  \frac{dQ_t}{dt} = \hbar \, \mathfrak{m}^{-1} \, \im \frac{(\psi,
  \nabla \psi)}{(\psi, \psi)} (Q_t)
\end{equation}
where $\mathfrak{m}$ is the diagonal matrix with the masses as
entries, each mass $m_k$ appearing 3 times. That is, as soon as
$\mathfrak{m}$ is given, the information \z{about} which directions in
$\RRR^{3N}$ correspond to the single factors $\RRR^3$ becomes
irrelevant for defining the dynamics of $Q_t$.  Eq.~\eqref{Qlawm}
would as well define a dynamics on any Euclidean vector space
$\Euclideanspace$ of finite dimension, given a wave function $\psi$ on
$\Euclideanspace$ and a positive-definite symmetric endomorphism
$\mathfrak{m}: \Euclideanspace \to \Euclideanspace$.

It will be convenient to include the mass matrix $\mathfrak{m}$ in the
metric $g_{ab}$ of $\Euclideanspace$,
\begin{equation}\label{gm1}
  g_{ab} = \sum_{c=1}^{\dim \Euclideanspace} g'_{ac} \mathfrak{m}^c_b\,,
\end{equation}
where $g'_{ab}$ is the metric of $\Euclideanspace$ before the inclusion of
masses, and indices $a,b,c$ run through the dimensions of
$\Euclideanspace$. In the standard example of $\RRR^{3N}$, this \z{amounts
to introducing} the metric
\begin{equation}\label{gm2}
  g_{i\alpha,j\beta} = m_i \delta_{ij} \delta_{\alpha \beta}\,,
\end{equation}
where $i,j = 1, \ldots, N$ and $\alpha, \beta = 1,2,3$ (and the index
$i$ \z{occurring}  twice on the right is not summed over). \z{With $\nabla$
then defined using $g$ instead of $g'$},
\eqref{Qlawm} becomes
\begin{equation}\label{Qlaw}
  \frac{dQ_t}{dt} = \hbar \, \im \frac{(\psi, \nabla \psi)}{(\psi,
  \psi)} (Q_t) \,. 
\end{equation}
(Note that in order to turn the covector given by the differential of
$\psi$ into a vector, one uses $g^{ab}$.) 

Similarly, the Schr\"odinger equation \eqref{e:sch01} can then be
written
\begin{equation}\label{e:sch01a}
  i\hbar \frac{\partial \psi}{\partial t} = - \tfrac{\hbar^2}{2} \gD
  \psi + V \psi\,,
\end{equation}
where $\gD$, the Laplacian on $\Euclideanspace$, is to be understood
as the metric trace of the second derivatives,
\begin{equation}
  \gD = g^{ab} \partial_a \partial_b
\end{equation}
(in abstract-index notation with sum convention). Thus, also the
Schr\"odinger equation is well-defined on a Euclidean space
$\Euclideanspace$, or, in other words, it is independent of the
product structure of $\RRR^{3N} = (\RRR^3)^N$.

\subsection{Riemannian Manifolds} 
\label{s:bbun}

In order to transfer \eqref{Qlaw} and \eqref{e:sch01a} to Riemannian
manifolds, we need only replace $\Euclideanspace$ by the tangent space
$T_{Q(t)} \Q$.  In this subsection, the wave functions we consider are
$\sspa$-valued functions on \gencon, with $\sspa$ a Hermitian vector
space.

We begin \z{by}  recalling the definitions of the gradient and the
Laplacian on Riemannian manifolds.  By the gradient $\nabla f$ of a
function $f: \Q \to \RRR$ we mean the tangent vector field on $\Q$
metrically equivalent (by ``raising the index'') to the 1-form $df$,
the differential of $f$. For a function $\psi: \Q \to \sspa$, the
differential $d\psi$ is a $\sspa$-valued 1-form, and thus $\nabla \psi
(q) \in \CCC T_q \Q \otimes \sspa$, where $\CCC T_q \gencon$ denotes
the complexified tangent space at $q$, and the tensor product
$\otimes$ is, as always in the following, over the complex numbers.
The Laplacian $\gD f$ of a function $f$ is defined to be the
divergence of $\nabla f$, where the divergence of a vector field $X$
is defined by
\begin{equation}
  \mathrm{div}\, X = D_a X^a
\end{equation}
with $D$ the (standard) covariant derivative operator, corresponding
to the Levi-Civita connection on the tangent bundle of $\Q$ arising
from the metric $g$.  Since $D g =0$, we can write
\begin{equation}
  \gD f = g^{ab} D_a D_b f \,,
\end{equation}
where the second $D$, the one which is applied first, actually does
not make use of the Levi-Civita connection. In other words, the
Laplacian is the metric trace of the second (covariant) derivative.
Another equivalent \z{definition is $\gD f = *d*d f$} 
where $d$ is the exterior derivative of differential forms and $*$ \z{is} the
Hodge star operator (see, e.g., \cite{EL83}).\footnote{The Hodge
  operator $*$ depends on the orientation of $\Q$ in such a way that a
  change of orientation changes the sign of the result. Thus, $*$ does
  not exist if $\Q$ is not orientable. However, it exists locally for
  any chosen local orientation, and since the Laplacian contains two
  Hodge operators, it is not affected by the sign ambiguity.} For
$\sspa$-valued functions $\psi$ the Laplacian $\gD\psi$ is defined
correspondingly as the divergence of the ``$\sspa$-valued vector
field'' $\nabla \psi$, or equivalently by
\begin{equation}
  \gD \psi = g^{ab} D_a D_b \psi 
\end{equation}
\z{or by $\gD \psi = *d*d \psi$}, using the obvious
extension of the exterior derivative to $\sspa$-valued differential
forms.


The time evolution of the state $(Q_t, \psi_t)$ is simply given by the
same formal equations as \eqref{Qlaw} and \eqref{e:sch01a} with the
appropriate interpretation of $\nabla$ and $\gD$. We give the
equations for future reference:
\begin{subequations} \label{ie:re}
\begin{align}
  \sdud{Q_t}{t} &= v^{\psi_t}(Q_t)\label{ie:rbe} \\
  i \hb \dud{\psi_t}{t}  &= - \tfrac{\hb^2}{2} \gD \psi_t
  + V \psi_t
  \, ,\label{ie:rse}
\end{align}
\end{subequations}
where the Bohmian velocity vector field $v^{\psi}$ associated to the
\wf\ $\psi$ is
\begin{equation} \label{ie:rbv}
  v^{\psi} \defi \hb\, \im \frac{\inpr{\psi}{\nabla
  \psi}}{\inpr{\psi}{\psi}}. 
\end{equation}

Thus, given $\gencon$, $\sspa$, and $V$, we have specified a Bohmian
dynamics, the \emph{immediate Bohmian dynamics}.\footnote{\z{The question
arises whether these equations possess unique solutions, for all times or
at least for short times. For some Riemannian manifolds $\Q$ this may
require the introduction of boundary conditions. Since the existence
question is mathematically demanding and not our concern here, we} make
only a few remarks: For a discussion of the \z{existence} question of
Bohmian trajectories in $\RRR^{3N}$, see \cite{BDGPZ95,TT04}.  \z{The
existence of} the evolution of the wave function amounts to defining the
Hamiltonian $H$ as a self-adjoint operator, i.e., as a self-adjoint
extension of $H^0 = \rawH$, with $H^0$ defined on $C_0^\infty (\Q,\sspa)$,
the space of smooth $\sspa$-valued functions with compact support.  As far
as we know, it is not known in all cases whether a self-adjoint extension
exists, and, when so, how many exist, and what the physical meaning of the
different extensions is when there is more than one. When several
extensions exist, they must perhaps be regarded as different possible
Bohmian dynamics on $\Q$, and thus as further possibilities, not captured
in the classes $\class_0, \class_1, \class_2$ considered in this paper. We
shall not pursue this idea further, and shall reason instead in terms of
the ``formal'' dynamics.}  We introduce the notation $\class_0(\Q,\sspa,V)$
for the set containing just this one dynamics. We also write
$\class_0(\Q,V)$ for $\class_0(\Q,\CCC,V)$. (A rigorous definition of what
is meant here by a ``dynamics,'' \z{avoiding the question of the existence
of solutions,} is given in \cite{topid1B}.  For now, we \z{simply proceed
as if we have the global existence of solutions and say (a bit vaguely)}
that a ``dynamics'' is defined by a set of wave functions, in this case
$C_0^\infty (\Q,\sspa)$, and for every wave function $\psi$ a set
$\set^\psi$ of trajectories in $\Q$, in this case the solutions of
\eqref{ie:rbe}, together with a probability distribution $\rho^\psi$ on
$\set^\psi$, in this case given by
\begin{equation}\label{rhopsisetpsi}
  \rho^\psi(dQ) = \bigl( \psi_t(Q_t), \psi_t(Q_t) \bigr) \, dQ_t \,.
\end{equation}
Due to equivariance, \eqref{rhopsisetpsi} is independent of $t$.)

\subsection{An Example}

An important \z{case is that}  of several particles moving in \z{a} 
Riemannian manifold $M$, a \z{possibly} curved physical space. Then the
configuration space for $N$ distinguished particles is $\gencon \defi
M^{N}$.  Let the masses of the particles be $m_i$ and the metric of
$M$ be $g$. Then the relevant metric on $M^N$, the analogue of
\eqref{gm1} and \eqref{gm2} acting on the tangent space $T_{(\canq_1,
\ldots, \canq_N)} M^N = \bigoplus\limts_{i=1}^{N} T_{\canq_i} M$, is
\[
  g^N(v_1 \oplus \cdots \oplus v_N, w_1 \oplus \cdots \oplus w_N )
  \defi \sum\limts_{i=1}^N m_i g(v_i,w_i). 
\]
Using $g^N$ allows us to write \eqref{ie:rbv} and \eqref{ie:re}
instead of the equivalent equations
\begin{equation}\label{e:bohm02}
  \sdud{\boldsymbol{Q}_{k}}{t} = \qmu{k} \im
  \frac {\inpr{\psi}{\nabla_{k} \psi}}{ \inpr{\psi}{\psi}}
  (\boldsymbol{Q}_1,\ldots, \boldsymbol{Q}_N),
  \quad  k= 1,\ldots,N
\end{equation}
\begin{equation}\label{e:sch02}
  \seq{N},
\end{equation}
where $\boldsymbol{Q}_k$, the $k^{th}$ component of $Q$, lies in $M$, and
$\nabla_k$ and $\gD_k$ are the gradient and the Laplacian with respect to
$g$, acting on the $k^{th}$ factor of $M^N$. We take $\sspa=\CCC$. Observe
that \eqref{e:bohm} and \eqref{e:sch01} are special cases, corresponding to
Euclidean space $M=\rvarn{3}$, of \eqref{e:bohm02} and \eqref{e:sch02}.

The configuration space of $N$ identical particles in $M$ is
\begin{equation}
  {}^N M := \{S| S \subseteq M, |S| = N\} \,,
\end{equation}
which inherits a Riemannian metric from $M$, see \cite{topid2}.

\subsection{Vector Bundles}
\label{sec:immedvb}

Even more generally, we can consider a Bohmian dynamics for \wf s
taking values in a complex vector bundle $E$ over the Riemannian
manifold \gencon. That is, the value space then depends on the
configuration, and \wf s become sections of the vector
bundle.\footnote{Recall that a \emph{section} (also known as
  \emph{cross-section}) of $E$ is a map $\psi:\gencon \to E$ such that
  $\psi(\baseq) \in E_{\baseq}$, i.e.\ it maps a point \baseq\ of
  \gencon\ to an element of the vector fiber over \baseq. For example,
  a vector field on a manifold $M$ is a section of the tangent bundle
  $TM$.}

Such a case occurs for identical particles with spin $s$, where the
bundle $E$ of spin spaces over the configuration space $\Q = \nrd$
defined in \eqref{nrddef} consists of the $(2s+1)^N$-dimensional
spaces
\begin{equation}\label{spinbundle}
  E_q = \bigotimes_{\canq \in q} \cmplx^{2s+1} \,, \quad q \in \Q\,. 
\end{equation}
For a detailed discussion of this bundle, of why this is the right
bundle, and of the notion of a tensor product over an arbitrary index
set, see  \cite{topid2}.  Vector bundles also occur for
particles with spin in a curved physical space. In addition to their
physical relevance, bundles are a natural mathematical generalization
of our previous setting involving wave functions defined on manifolds.
Finally,  the approaches we use in \cite{topid1B,topid1D} for
suggesting natural classes of Bohmian dynamics are based on
considerations concerning vector bundles (even for spinless
particles).

We introduce some notation and terminology.  $C^{\infty} (E)$ will
denote the set of smooth sections while $C^{\infty}_0 (E)$ will be the
set of smooth sections with compact support. 

\begin{defn}
  A \emph{\herm\ vector bundle}, \z{or}  \emph{\herm\ bundle}, is a
  finite-dimensional complex vector bundle $E$ with a connection and a
  positive-definite, Hermitian local inner product
  $(\,\cdot\,,\,\cdot\,)_q$ on $E_q$ which is parallel.
\end{defn}

Recall that a connection defines (and is defined by) a notion of parallel
transport of vectors in $E_q$ along \z{curves} $\opencurve$ \z{in $\Q$}
from $q$ to $r$, \z{given by linear mappings} $P_\opencurve: E_q \to
E_r$. A section $\psi$ of $E$ is \emph{parallel} if always $P_\opencurve
\psi(q) = \psi(r)$. If $\opencurve$ is a loop, $q=r$, the mapping
$P_\opencurve$ is called the \emph{holonomy endomorphism}
$\holonomy_\opencurve$ of $E_q$ associated with $\opencurve$.  A connection
also defines (and is defined by) a covariant derivative operator $D$, which
allows us to form the derivative $D\psi$ of a section $\psi$ of $E$. A
section $\psi$ is parallel if and only if $D\psi =0$. A bundle with
connection is called \emph{flat} if all holonomies of contractible loops
are trivial, i.e., the identity endomorphism (this is the case if and only
if the curvature of the connection vanishes everywhere).

Parallelity of the local inner product means that parallel transport
preserves inner products; equivalently, $D(\psi,\phi) = (D\psi,\phi) +
(\psi, D\phi)$ for all $\psi,\phi \in C^\infty(E)$.  It follows in
particular that holonomy endomorphisms are always unitary.

Our bundle, the one of which $\psi$ is a section, will always be a
\herm\ bundle. Note that since a \herm\ bundle consists of a vector
bundle and a connection, it can be nontrivial even if the vector
bundle is trivial: namely, if the connection is nontrivial. The
\emph{trivial \herm\ bundle} $\Q \times \sspa$, in contrast, consists
of the trivial vector bundle with the trivial connection, whose
parallel transport $P_\opencurve$ is always the identity on $\sspa$.
The \z{case of a $W$-valued function}  $\psi : \Q \to \sspa$ corresponds to the trivial \herm\ 
bundle $\Q \times \sspa$.

The global inner product on the Hilbert space of \wf s is the
local inner product integrated against the Riemannian volume measure
associated with the metric $g$,
\[
  \langle \phi, \psi \rangle = \int_{\gencon} dq \, (\phi(q),
  \psi(q))\,. 
\]
The Hilbert space equipped with this inner product, denoted
$L^2(\gencon,E)$, contains the square-integrable, measurable (not
necessarily smooth) sections of $E$ modulo equality almost everywhere. 
In an obvious sense, $C_0^\infty(E) \subseteq L^2(\gencon,E)$. 

The covariant derivative $D\psi$ of a section $\psi$ is an ``$E$-valued
1-form,'' i.e., a section of $\CCC T\gencon^* \otimes E$ (with $T\Q^*$ the
cotangent bundle), while we write $\nabla \psi$ for the section of $\CCC
T\gencon \otimes E$ metrically equivalent to $D\psi$.  To define the
covariant derivative of $D\psi$, one uses the connection on $\CCC
T\gencon^* \otimes E$ that arises in an obvious way from the Levi-Civita
connection on $\CCC T \gencon^*$ and the given connection on $E$, with the
defining property $D_{\CCC T \gencon^* \otimes E} (\omega \otimes \psi) =
(D_{\CCC T \gencon^*} \omega) \otimes \psi + \omega \otimes (D_E \psi)$ for
every 1-form $\omega$ and every section $\psi$ of $E$. We take as the
\z{Laplacian $\gD \psi$} of $\psi$  the (Riemannian)
metric trace of the second covariant derivative of $\psi$,
\begin{equation}\label{gDdef}
  \gD \psi = g^{ab} D_a D_b \psi\,, 
\end{equation}
where the second $D$, the one which is applied first, is the covariant
derivative on $E$, and the first $D$ is the covariant derivative on $\CCC
T\Q^* \otimes E$.\footnote{While this is the natural definition
  of the Laplacian of a section of a \herm\ bundle, we note that for
  differential $p$-forms with $p \geq 1$ there are two inequivalent
  natural definitions of the Laplacian: one is $\gD = -(d^*d + dd^*)$
  (sometimes called the de Rham Laplacian, with $d^* = (-1)^{(\dim \Q)
    (p+1)+1} *d*$ on $p$-forms \cite[p.~9]{EL83}), the other is
  \eqref{gDdef} for $E = \CCC \Lambda^p T\Q^*$ (sometimes called the
  Bochner Laplacian). They differ by a curvature term given by the
  Weitzenb\"ock formula \cite[p.~11]{EL83}.}
Again, an equivalent definition is $\gD \psi = *d*d \psi$, using the
obvious extension (based on the connection of $E$) of the exterior
derivative to $E$-valued differential forms, i.e., sections of $\CCC
\Lambda^p T \Q^* \otimes E$.

The potential $V$ is now a self-adjoint section of the endomorphism
bundle $E \otimes E^*$ acting on the vector bundle's fibers.

The equations defining the Bohmian dynamics are the same as before.
Explicitly, we define $v^{\psi}$, the Bohmian velocity vector field
associated \z{with}  a \wf\ $\psi$, by
\begin{equation} \label{ie:vbv}
  v^{\psi} \defi \hb\, \im \frac{\inpr{\psi}{\nabla
  \psi}}{\inpr{\psi}{\psi}}. 
\end{equation}
The time evolution of
the state $(Q_t, \psi_t)$ is given by
\begin{subequations} \label{ie:ve}
\begin{align}
  \sdud{Q_t}{t} &= v^{\psi_t}(Q_t) \label{ie:vbe} \\
  i \hb \dud{\psi_t}{t} &= - \tfrac{\hb^2}{2} \gD \psi_t  + V \psi_t
  \label{ie:vse}
\end{align}
\end{subequations}
\z{The class} $\class_0(\Q, E, V)$ contains just this one
dynamics, defined by \eqref{ie:vbv} and \eqref{ie:ve}.  This agrees
with the definition of $\class_0(\Q, \sspa, V)$ given in
\Sect{\ref{s:bbun}} in the sense that $\class_0(\Q, \sspa, V) =
\class_0 (\Q, E, V)$ when $E$ \z{is the}  trivial bundle $\Q
\times \sspa$. 

Equivariance of the distribution $\rho = (\psi,\psi)$ is (on a formal
level) obtained from the equations
\begin{equation}\label{continuityv2}
  \dud{\rho_t}{t} =- \nabla \cdot (\rho_t v^{\psi_t}) 
\end{equation}
for the distribution $\rho_t$ of the configuration $Q_t$ and
\begin{equation}\label{dpsi2dtv2}
  \dud{|\psi_t|^2}{t} =- \nabla \cdot (|\psi_t|^2 v^{\psi_t}),
\end{equation}
which follow, \z{since $(\,\cdot\, ,\,\cdot\,)$ is parallel,} from
\eqref{ie:vbv} and \eqref{ie:ve} \z{just as \eqref{continuity} and
\eqref{dpsi2dt} follow from} \eqref{e:bohm} and \eqref{e:sch01}.

\section{Scalar Periodic Wave Functions on the Covering Space}
\label{sec:covering}

We introduce now the Bohmian dynamics belonging to the class that we denote
$\class_1$; in \Sect{\ref{s:periodic2}} we introduce the dynamics of class
$\class_2$.  To this end, we will consider wave functions on the universal
covering space of $\Q$. This idea is rather standard in the literature on
quantum mechanics in multiply-connected spaces \cite{DL71, D72, LM77,
Mor92, HM96}.  However, the standard treatment lacks the precise
justification that one can provide in Bohmian mechanics. Moreover, the
complete classification of the possibilities that we give in
\Sect{\ref{s:periodic2}} includes some, corresponding to what we call
\emph{holonomy-twisted representations} of $\fund{\Q}$, that until recently
\cite{topid0} had not been considered.  The possibilities considered so far
correspond to unitary representations of $\fund{\Q}$ on the value space of
the wave function.  Each possibility has locally the same Hamiltonian
$\rawH$, with the same potential $V$, and each possibility is equally well
defined and equally reasonable. In this section all \wf s will be
complex-valued; in \Sect{\ref{s:periodic2}} we consider wave functions with
higher-dimensional value spaces.

\subsection{The Circle, for Example} 
\label{is:circ}

Let us start with the \z{configuration space $\gencon = S^1$}, the
circle. This space is multiply connected since only those loops that
surround the circle as many times clockwise as counterclockwise \z{can be
shrunk to a point}. It is convenient to write the wave function $\psi: S^1
\to \cmplx$ as a function $\hat{\psi}(\theta)$ of the angle coordinate,
with $\hat{\psi}: \re \to \cmplx$ a $2\pi$-periodic function. From
$\hat{\psi}$ one obtains
\begin{equation}\label{vhatdefcircle}
  \hat{v}^{\hat{\psi}} = \hbar \, \im \, \frac{
  \nabla \hat{\psi}}{ \hat{\psi} }
\end{equation}
as a $2\pi$-periodic function of $\theta$. The relevant observation is
that for \eqref{vhatdefcircle} to be $2\pi$-periodic, it is
(sufficient but) not necessary that $\hat{\psi}$ be
$2\pi$-periodic. It would be sufficient as well to have a $\hat{\psi}$
that is merely periodic up to a phase shift,
\begin{equation} \label{ie:sper}
\hat {\psi} (\theta + 2\pi) = \gamma \hat {\psi} (\theta),
\end{equation}
where $\gamma$ is a complex constant of modulus one, called a
\emph{topological phase factor}.

Another way of viewing this is to write $\psi$ in the polar form
$Re^{iS/\hb}$, where $R\geq 0$ and the phase $S$ is real, and find
that the Bohmian velocity \eqref{ie:rbv} is given by $v^{\psi} =
\nabla S$. If we view the phase $S$ as a function $\hat{S}(\theta)$ of
the angle coordinate, we see that $\nabla \hat{S}$ will be
$2\pi$-periodic if
\begin{equation}\label{Speriodic}
  \hat{S}(\theta + 2\pi) = \hat{S}(\theta) + \beta
\end{equation}
for some constant $\beta \in \re$. This corresponds to \eqref{ie:sper}
\z{with}  $\gamma:= e^{i\beta/\hbar}$. 

Since $\hat{v}^{\hat{\psi}}$ is $2\pi$-periodic, it makes sense to
write the equation of motion
\begin{equation}\label{e:bohm03}
  \frac{dQ_t}{dt} = \hat{v}^{\hat{\psi}}(\theta(Q_t)) = \hbar \, \im \,
  \frac{ \nabla \hat{\psi}}{
  \hat{\psi} }(\theta(Q_t))
\end{equation}
where $\theta(Q_t)$ is any of the values of the angle coordinate that
one can associate with $Q_t$. If we let $\hat{\psi}$ evolve by the
Schr\"odinger equation on the real line with a $2\pi$-periodic
potential $V$,
\begin{equation}\label{e:sch03}
  i\hbar\frac{\partial \hat{\psi}_t}{\partial t} = -\tfrac{\hbar^2}{2}
  \gD \hat{\psi}_t + V\hat{\psi}_t,
\end{equation}
then the periodicity condition \eqref{ie:sper} is preserved by the
evolution, thanks to the linearity of the Schr\"odinger equation.
Thus, for any fixed complex $\gamma$ of modulus one, \eqref{e:bohm03},
\eqref{e:sch03}, and \eqref{ie:sper} together define a Bohmian
dynamics, \z{just}  as \eqref{ie:rbv} and \eqref{ie:re} do.
This dynamics permits as many different wave functions as the one
defined by \eqref{ie:rbv} and \eqref{ie:re}, which corresponds to
$\gamma =1$.  

\z{Since} $|\gamma| =1$, \z{so that $|\hat{\psi}|^2$ is $2\pi$-periodic,}
this theory also has an equivariant probability distribution on the circle,
with density $\rho = |\hat{\psi}|^2$. This is the reason why we restrict
the possibilities \z{to $\gamma$ of modulus 1:} otherwise we lose
equivariance. (The trajectories may still exist \z{globally} even if
$|\gamma| \neq 1$.)

We summarize the results of our reasoning. 

\begin{assertion}\label{a:circle}
  For each potential $V$ and each complex number $\gamma$ of modulus
  one, there is a Bohmian dynamics on the circle, defined by
  \eqref{ie:sper}, \eqref{e:bohm03}, and \eqref{e:sch03}. 
\end{assertion}

According to the notation that we will define later, these dynamics
form the class $\class_1(S^1,V)$, a one-parameter class parametrized
by $\gamma$. What are the physical factors that determine which
$\gamma$ is to be used? It depends. The following subsection provides
a concrete example.

\subsection{Relation to the Aharonov--Bohm Effect} 
\label{s:ab}

The additional possibilities associated with nontrivial phase shifts
$\gamma$ occur in physics even in the case of the circle. We describe
here a simplified version of the Aharonov--Bohm effect
\cite{AB59,Gri94,PT89}.

Consider a single particle \z{confined}  to a loop in 3-space.  \z{Suppose
there is}   a magnetic field $\magnetic$ that vanishes at
every point of the loop, \z{but with} field lines \z{that}  pass through the
interior of the loop. \z{Thus,}  if $D$ is a 2-dimensional surface
bounded by the loop, there may be a nonzero flux of the magnetic field
across $D$,
\begin{equation}\label{Phidef}
  \Phi := \int_{D}  \magnetic \cdot \boldsymbol{n} \, dA,
\end{equation}
where $dA$ is the area element and $\boldsymbol{n}$ is the unit
normal on the surface $D$. (Note that by Maxwell's equation $\nabla
\cdot \magnetic =0$ and the Ostrogradski--Gauss integral formula, the
value of $\Phi$ does not depend on the particular choice of the
surface $D$.)

The appropriate quantum or Bohmian theory on $S^1$ corresponds, in the
sense of Assertion~\ref{a:circle}, to the phase factor
\begin{equation}\label{ABphase}
  \gamma = e^{-ie\Phi/\hbar}\,,
\end{equation}
where $e$ in the exponent is the charge of the particle, provided that
the orientation of the loop and the surface agree in the sense that
the direction of increasing $\theta$ and the direction of
$\boldsymbol{n}$ satisfy a right-hand-rule. A justification of this
description will be given in \Sect{\ref{s:ab2}}.

From this example we conclude that the dynamics of class $\class_1
\setminus \class_0$ \emph{do actually occur} in a universe governed by
Bohmian mechanics as the effective dynamics in a restricted
configuration space.

\subsection{Notation and Relevant Facts Concerning Covering Spaces}
\label{sec:notation}

In the remainder of this section, we generalize the \z{considerations}  of
\Sect{\ref{is:circ}} to arbitrary $\Q$.  The relevant notation is
summarized by the table:
\begin{center}
\begin{tabular}{ccc}
& Generic & Simplest example \\
configuration space & \gencon & $S^1$ \\
universal covering space & \covspa & \rvarn{} \\
points in \gencon, \covspa & \baseq, \covq & $e^{i\theta}$, $\theta$ \\
projection map & $\proj : \covspa \to \gencon$ & $e^{i \cdot}:
  \rvarn{}\to S^1$\\
covering fiber over \baseq, $e^{i\theta}$ & $\proj^{-1} (\baseq)$ &
  $\{\theta + 2\pi k\,|\, k \in \mathbb{Z}\}$ \\
fundamental group of \gencon & \fund{\gencon} & $\mathbb{Z}$ \\
covering transformation & $\deckt: \covspa \to \covspa$
   & $\sigma_k: \theta \mapsto \theta+2\pi k$ \\
covering group & \deckg & $\{\sigma_k \,|\, k \in \mathbb{Z}\}$ \\
\z{character of fundamental group}  & $\concav_{\deckt}$ & $\concav_{k} =
  \concav^{k}$  \\
bundle, lifted bundle & $E$, $\wh{E}$ & $S^1\times \cmplx$,
  $\rvarn{}\times \cmplx$ \\
\end{tabular}
\end{center}

Again, \gencon\ \z{is} a Riemannian manifold with metric $g$, \z{with
  universal covering space denoted by \covspa.}  Recall that the universal
  covering space is, by definition, a simply connected space, endowed with
  a covering map (a local diffeomorphism) $\proj: \covspa \to \gencon$,
  also called the projection. The \emph{\fiber} for $\baseq\ \in \gencon$
  is the set $\proj^{-1}(\baseq)$ of points in \covspa\ that project to
  \baseq\ under $\proj$. Every function or vector field on \gencon\ can be
  lifted to a function, respectively vector field, on \covspa. The
  functions and vector fields on \covspa\ arising in this way are called
  \emph{projectable}. A function $f: \covspa \to \cmplx$ is projectable
  if and only if $f(\covq) = f(\covr)$ whenever $\proj(\covq) =
  \proj(\covr)$. In that case it is the lift of $\tilde{f} : \gencon \to
  \cmplx$ given by $\tilde{f}(\proj (\covq)) \defi f(\covq )$, called the
  projection of $f$.  A vector field $w$ on \covspa\ is projectable if and
  only if, whenever $\proj(\covq) = \proj(\hat r )$, $\proj^* w (\covq) =
  \proj^* w (\covr)$ where $\proj^*$ is the (push-forward) action of
  $\proj$ on tangent vectors.

We shall always take \covspa\ to be endowed with the lifted metric
$\hat{g}$, which makes \covspa\ a Riemannian manifold as well and
assures that $\proj$ is a local isometry. As a consequence, if
$\hat{f}$ is the lift of the function $f$, $\gD_{\covspa} \hat{f} =
\widehat{\gD_\gencon f}$. 

A \emph{covering transformation} is an isometry $\deckt$ mapping the
covering space to itself which preserves the \fiber s, $\proj \circ \deckt
= \proj$.  The group of such transformations is the \emph{covering group}
and is denoted by \deckg. It acts freely and transitively on every covering
fiber, i.e., for every $\covq$ and $\hat r$ in the same fiber there is
precisely one $\deckt$ such that $\hat r = \deckt \covq$. As a consequence,
projectability of the vector field $w$ on \covspa\ is equivalent to \z{the
condition that} $w(\deckt \covq) = \deckt^* w(\covq)$ for all $\covq \in
\covspa$ and all $\deckt \in \deckg$.

The \emph{fundamental group at a point} \baseq, denoted by \fund{\gencon,
\baseq}, is the set of equivalence classes of closed loops through \baseq,
where the equivalence relation is that of homotopy, i.e.\ smoothly
deforming one curve into the other.  The product in \z{this} group is
concatenation; more precisely, \z{$\sigma \tau$ corresponds to the loop
obtained by} first \z{following} $\tau$ and then \z{following}
$\sigma$. (This is in contrast to the common definition of the product,
\z{with}  the opposite order. We do it this way as it seems more natural
\z{for}  parallel transport.)  The fundamental groups at different
points are isomorphic to each other as well as to the covering group, but
the isomorphisms are not canonical. However, for every given $\covq \in
\covspa$ there is a canonical isomorphism $\varphi_{\covq}: \deckg \to
\fund{\gencon, \proj(\covq)}$; for different choices of $\covq$ in the same
fiber, \z{the different $\varphi_{\covq}$'s are conjugate:} they are related by $\varphi_{\deckt \covq} (\tau) =
\varphi_{\covq}( \deckt^{-1} \tau \deckt) = \varphi_{\covq}(\deckt)^{-1} \,
\varphi_{\covq}(\tau) \, \varphi_{\covq}(\deckt)$.  By \emph{the
fundamental group of} $\gencon$, written \fund{\gencon}, we shall mean any
one of the fundamental groups \fund{\gencon, \baseq}.

A \emph{character} of a group $G$ is a unitary, 1-dimensional
representation of $G$, i.e., a homomorphism $G \to U(1)$ where $U(1)$
is the multiplicative group of the complex numbers of modulus one. The
characters of $G$ form a group denoted \z{by} $G^*$.

\subsection{Scalar Periodic Wave Functions}
\label{sec:scalarperiodic}

The motion of the configuration $Q_t$ in \gencon\ is determined by a
velocity vector field $v_t$ on \gencon, which may arise from a wave
function $\psi$ not on \gencon\ but instead on \covspa\ in the
following way. 

Suppose we are given a map $\gamma: \deckg \to \cmplx$, and suppose
that a wave function $\psi: \covspa \to \cmplx$ satisfies the
\emph{periodicity condition associated with the topological factors
  $\gamma$}, i.e.,
\begin{equation} \label{e:percon}
  \psi (\deckt \covq) = \concov \psi(\covq)
\end{equation}
for every $\covq \in \covspa$ and $\deckt \in \deckg$. (We no longer
put the hat $\hat{\ }$ on top of $\psi$ that served for emphasizing
that $\psi$ lives on the covering space.) For \eqref{e:percon} to be
possible for a $\psi$ that does not identically vanish, $\gamma$ must
be a representation of the covering group, as was first emphasized in
\cite{D72}. To see this, let $\deck1$, $\deck2 \in \deckg$. Then we
have the following equalities
\begin{equation}\label{e:cccargument}
\conc{\deck1 \deck2} \psi (\covq)
= \psi(\deck1 \deck2 \covq)
= \conc{\deck1} \psi(\deck2 \covq)
= \conc{\deck1}\conc{\deck2} \psi(\covq). 
\end{equation}
We \z{thus obtain}  the fundamental relation
\begin{equation} \label{e:ccc}
  \conc{\deck1 \deck2} = \conc{\deck1}\conc{\deck2},
\end{equation}
establishing \z{(since $\gamma_{\id} =1$)}  that $\gamma$
is a representation. 

The 1-dimensional representations of the covering group are, \z{via the
canonical isomorphisms $\varphi_{\covq}: \deckg \to \fund{\gencon, q},\
\covq\in \proj^{-1}(\baseq)$,} in canonical correspondence with the
1-dimensional representations of any fundamental group \fund{\gencon, q}:
\z{The different} isomorphisms \z{$\varphi_{\covq},\ \covq\in
\proj^{-1}(\baseq)$,} will \z{transform a representation} of
\fund{\gencon, q} into \z{representations of $\deckg$ that are
conjugate. But the} 1-dimensional representations are homomorphisms to the
\emph{abelian} multiplicative group of $\cmplx$ and \z{are} thus invariant
under conjugation.

{}From \eqref{e:percon} it follows that $\nabla \psi(\deckt \covq) =
\gamma_\deckt \, \deckt^* \nabla \psi(\covq)$, where $\deckt^*$ is the
(push-forward) action of $\deckt$ on tangent vectors, using that
$\deckt$ is an isometry. Thus, the velocity field $\hat{v}^{\psi}$ on
\covspa\ associated with $\psi$ according to
\begin{equation}\label{vhatdef}
  \hat{v}^\psi (\covq) := \hbar \, \im \, \frac{ \nabla
  \psi}{ \psi} (\covq)
\end{equation}
is projectable, i.e.,
\begin{equation}\label{vhatprojectable}
  \hat{v}^\psi (\deckt\covq) = \deckt^* \hat{v}^\psi (\covq),
\end{equation}
and therefore gives rise to a velocity field $v^\psi$ on
\gencon,
\begin{equation}
  v^\psi(q) = \proj^* \, \hat{v}^\psi (\covq)
\end{equation}
where $\covq$ is an arbitrary element of $\proj^{-1}(q)$. 

If we let $\psi$ evolve according to the Schr\"odinger equation on \covspa,
\begin{equation}\label{e:sch04}
  i\hbar \frac{\partial \psi}{\partial t}(\covq) = - \tfrac{\hbar^2}{2} \gD
  \psi(\covq) + \widehat{V}(\covq) \psi(\covq)
\end{equation}
with $\widehat{V}$ the lift of the potential $V$ on $\gencon$, then
the periodicity condition \eqref{e:percon} is preserved by the
evolution, since, according to
\begin{equation}
  i\hbar\frac{\partial \psi}{\partial t}(\deckt \covq)
  \stackrel{\eqref{e:sch04}}{=} -\tfrac{\hbar^2}{2} \gD \psi(\deckt
  \covq) + \widehat{V}(\deckt \covq) \psi(\deckt \covq) =
  -\tfrac{\hbar^2}{2} \gD \psi(\deckt \covq) + \widehat{V}(\covq)
  \psi(\deckt \covq)
\end{equation}
(note the different arguments in the potential), the functions $\psi
\circ \deckt$ and $\gamma_\deckt \psi$ satisfy the same evolution
equation \eqref{e:sch04} with, by \eqref{e:percon}, the same initial
condition, and thus coincide at all times.

\z{Therefore}  we can let the Bohmian configuration $Q_t$ move according
to $v^{\psi_t}$,
\begin{equation}\label{e:bohm04}
  \frac{dQ_t}{dt} = v^{\psi_t}(Q_t) = \hbar\, \proj^* \Bigl( \im\,
  \frac{ \nabla \psi}{
  \psi}\Bigr)(Q_t) = \hbar\, \proj^* \Bigl( \im\,
  \frac{ \nabla \psi}{
  \psi}\Big|_{\covq \in \proj^{-1}(Q_t)} \Bigr). 
\end{equation}
One can also view the motion in this way: Given $Q_0$, choose
$\widehat{Q}_0 \in \proj^{-1}(Q_0)$, let $\widehat{Q}_t$ move in
\covspa\ according to $\hat{v}^{\psi_t}$, and set $Q_t =
\proj(\widehat{Q}_t)$. Then the motion of $Q_t$ is independent of the
choice of $\widehat{Q}_0$ in the fiber over $Q_0$, and obeys
\eqref{e:bohm04}. 

If, as we shall assume from now on, $|\gamma_\deckt|=1$ for all
$\deckt \in \deckg$, i.e., if $\gamma$ is a \emph{unitary}
representation (in $\cmplx$) or a \emph{character}, then the motion
\eqref{e:bohm04} also has an equivariant probability distribution,
namely
\begin{equation}\label{e:equi04}
  \rho(q) = |\psi(\covq)|^2. 
\end{equation}
To see this, note that we have
\begin{equation}\label{projectablepsi2}
  |\psi(\deckt \covq)|^2 \stackrel{\eqref{e:percon}}{=}
  |\gamma_\deckt|^2 |\psi(\covq)|^2 = |\psi(\covq)|^2,
\end{equation}
so that the function $|\psi(\covq)|^2$ is projectable to a function on
\gencon\ which we call $|\psi|^2(q)$ in this paragraph. From
\eqref{e:sch04} we have \z{that} 
\[
  \frac{\partial |\psi_t(\covq)|^2}{\partial t} = - \nabla \cdot \Bigl(
  |\psi_t(\covq)|^2 \, \hat{v}^{\psi_t}(\covq) \Bigr)
\]
and, by projection, \z{that} 
\[
  \frac{\partial |\psi_t|^2(q)}{\partial t} = - \nabla \cdot \Bigl(
  |\psi_t|^2 (q)\, v^{\psi_t}(q) \Bigr),
\]
which coincides with the transport equation for a probability density
$\rho$ on \gencon,
\[
  \frac{\partial \rho_t(q)}{\partial t} = - \nabla \cdot \Bigl(
  \rho_t(q) \, v^{\psi_t}(q) \Bigr). 
\]
Hence, 
\begin{equation}
  \rho_t(q) = |\psi_t|^2(q)
\end{equation}
for all times if \z{it is} so initially; this is equivariance. 

This also makes clear that the relevant wave functions are those with
\begin{equation}
  \int_\gencon dq \, |\psi(\covq)|^2 = 1
\end{equation}
where the choice of $\covq \in \proj^{-1}(q)$ is arbitrary by
\eqref{projectablepsi2}. The relevant Hilbert space, which we denote
$L^2(\covspa,\gamma)$, thus \z{consists of}  the measurable functions $\psi$
on $\covspa$ (modulo changes on null sets) satisfying \eqref{e:percon}
with
\begin{equation}
  \int_\gencon dq \, |\psi(\covq)|^2 < \infty. 
\end{equation}
It is a Hilbert space with the scalar product
\begin{equation}
  \langle \phi,\psi \rangle = \int_\gencon dq \, \overline{\phi(\covq)}
  \, \psi(\covq). 
\end{equation}
Note that the value of the integrand at $q$ is independent of the choice of
$\covq \in \proj^{-1}(q)$ since, by \eqref{e:percon} and \z{the fact that}
$|\gamma_\deckt|=1$,
\[
  \overline{\phi(\deckt \covq)} \, \psi(\deckt \covq) =
  \overline{\gamma_\deckt \, \phi(\covq)} \, \gamma_\deckt \,
  \psi(\covq) = \overline{\phi(\covq)} \, \psi(\covq). 
\]

We summarize the results of our reasoning. 

\begin{assertion}\label{a:scalar}
  Given a Riemannian manifold $\gencon$ and a smooth function
  $V:\gencon \to \re$, there is a Bohmian dynamics in \gencon\ with
  potential $V$ for each character \concav\ of the fundamental group
  \fund{\gencon}; it is defined by \eqref{e:percon}, \eqref{e:sch04},
  and \eqref{e:bohm04}, where the wave function $\psi_t$ lies in
  $L^2(\covspa,\gamma)$ and has norm one. 
\end{assertion}

We define $\class_1(\Q,V)$ to be the class of Bohmian dynamics provided by
Assertion~\ref{a:scalar}.  It contains as many elements as there are
characters of $\fund{\Q}$ because different characters $\gamma' \neq
\gamma$ always define different dynamics; we give a proof of this fact in
\cite{topid1B}.\footnote{\label{fn:path}\z{But,} essentially, this is already
clear for the same reason as why one can, in the Aharonov--Bohm effect,
read off the phase shift from a shift in the interference pattern: If one
splits a wave packet, located at $q$, \z{into} two \z{pieces} and, say,
lets them move along curves $\opencurve_1$ and $\opencurve_2$ from $q$ to
$r$ \z{that are} such that the loop $\opencurve_1^{-1} \opencurve_2$ is
incontractible, then one obtains interference between the two packets at
$r$ in a way that depends on the phase shift associated with the loop
$\opencurve_1^{-1} \opencurve_2$.
  
  Interestingly, different characters can define the same dynamics
  when we consider, instead of complex-valued wave functions, sections
  of \herm\ bundles. \label{ft:charequivalence}Here is an example of
  such a nontrivial \herm\ bundle $E$: consider $\Q = \nrtre$, whose
  fundamental group is the permutation group $S_N$ with two
  characters, and the bundle $E = F \oplus B$ over $\nrtre$, where $F$
  is what we call the fermionic line bundle (the unique flat \herm\ 
  line bundle over $\nrtre$ whose holonomy representation is the
  alternating character) and $B = \nrtre \times \CCC$, \z{the trivial}   line
  bundle; see section 7.2 of \cite{topid2} for a discussion.}

\subsection{Remarks}
\label{is:rem}

\begin{enumerate}

\item Since the law of motion \eqref{e:bohm04} involves a derivative
  of $\psi$, the merely measurable functions in $L^2(\covspa,\gamma)$
  will of course not be \z{adequate} for defining trajectories. 
  However, we will leave aside the \z{question,}  from which dense
  subspace of $L^2(\covspa,\gamma)$ \z{should one}  choose $\psi$. 

\item In our example $\gencon = S^1$, the possible dynamics are
  precisely those mentioned in \Sect{\ref{is:circ}}. The covering
  group is isomorphic to $\mathbb{Z}$, and every homomorphism $\gamma$
  is of the form $\gamma_k = \gamma_1^k$.  \z{Thus}  a character is
  determined by  a complex \z{number $\gamma_1$} of modulus one; the
  periodicity condition \eqref{e:percon} reduces to \eqref{ie:sper}. 

\item For the trivial character $\gamma_\deckt =1$, we obtain the
  immediate dynamics, as defined by \eqref{ie:rbv} and \eqref{ie:re}.
  Thus, $\class_0(\Q,V) \subseteq \class_1(\Q,V)$.

\item\label{rem:idnospin} Another example, or application of Assertion~\ref{a:scalar}, is
  provided by identical particles without spin.  The natural
  configuration space $\nrd$ \z{for}  identical particles, defined in
  \eqref{nrddef}, has fundamental group $S_N$, the group of
  permutations of $N$ objects, which possesses two characters, the
  trivial character, $\gamma_\sigma =1$, and the alternating
  character, $\gamma_\sigma = \mathrm{sgn}(\sigma)= 1$ or $-1$
  depending on whether $\sigma \in S_N$ is an even or an odd
  permutation. As explained in detail in \cite{topid2}, the Bohmian
  dynamics associated with the trivial character is that of bosons,
  while the one associated with the alternating character is that of
  fermions.

\item \z{When} $|\gamma_\deckt| \neq 1$ for some $\deckt \in \deckg$, in
which \z{case} the equivariant distribution \eqref{e:equi04} is not
defined, one could think of obtaining instead an equivariant distribution
by setting \begin{equation}\label{altequi} \rho(q) = \sum_{\covq \in
\proj^{-1}(q)} |\psi(\covq)|^2 .  \end{equation} However, this ansatz does
not work for providing an equivariant distribution in \z{this case. Any
$\deckt$ for which $|\gamma_\deckt| \neq 1$} must be an element of infinite
order, since otherwise $\gamma_\deckt$ would have to be a root of unity.
\z{Thus} $\fund{\gencon}$ is infinite, and so is the covering fiber
$\proj^{-1}(q)$, which is in a canonical (given $\covq \in \proj^{-1}(q)$)
correspondence with $\fund{\gencon,q}$, and the sum on the right hand side
of \eqref{altequi} is divergent unless $\psi$ vanishes everywhere on this
covering fiber. (To see this, note that either $|\gamma_\deckt|>1$ or
$|\gamma_{\deckt^{-1}}|>1$ since $\gamma$ is a representation; without loss
of generality we suppose $|\gamma_\deckt|>1$. If $\psi(\covq) \neq 0$ for
some $\covq$, then already the sum over just the fiber elements $\deckt^k
\covq$, $k=1,2,3,\ldots$, is divergent, since by the periodicity condition
\eqref{e:percon}, $\sum_k |\psi(\deckt^k \covq)|^2 = \sum_k
|\gamma_{\deckt^k}|^2 |\psi(\covq)|^2 = |\psi(\covq)|^2 \sum_k
|\gamma_{\deckt}|^{2k} = \infty$.)

\item\label{rem:GRW}
  As stated already in Section \ref{sec:GRW}, topological factors can also
  be introduced into GRW theories, provided we start with a GRW theory of
  the following kind: Wave functions $\psi$ are functions on a Riemannian
  manifold $\Q$, and collapses according to \eqref{GRWcollapse} occur with
  rate \eqref{GRWrate} with collapse rate operators $\Lambda(\vx)$ (where
  $\vx$ is in for example $\RRR^3$) that are multiplication operators on
  configuration space:
\begin{equation}
   \Lambda(\vx) \, \psi (q) = f_{\vx}(q) \, \psi(q)\,.
\end{equation}
Then we may define a GRW theory with topological factor given by the character 
$\gamma$ of $\pi_1(\Q)$ by using wave functions $\psi$ on the covering 
space $\covspa$ satisfying the periodicity condition \eqref{e:percon} 
associated with $\gamma$, with collapse rate operators the lifted 
multiplication operators on $\covspa$:
\begin{equation}
   \Lambda(\vx) \, \psi(\hat{q}) = f_{\vx} (\pi(\hat{q})) \, 
\psi(\hat{q})\,.
\end{equation}
Collapse then maps periodic to periodic wave functions, with the same 
topological factor. Note, however, that the theory would work as well with 
aperiodic wave functions.

\setcounter{remi}{\theenumi}
\end{enumerate}

\section{The Aharonov--Bohm Effect}
\label{s:abs}

We now give a more detailed treatment of the Aharonov--Bohm effect in
the framework of Bohmian mechanics and its relation to topological
phase factors. In doing so, we repeat various standard considerations
on this topic.

\subsection{Derivation of the Topological Phase Factor}
\label{s:ab2}

To justify the dynamics described in \Sect{\ref{s:ab}}, we consider
a less idealized description of the Aharonov--Bohm effect. \z{Consider}  a particle
moving in $\RRR^3$ that cannot enter the solid cylinder
\begin{equation}
  \cylinder = \bigl\{ \canq = (q_1,q_2,q_3) \in \RRR^3 \,\big|\,
  q_1^2 + q_2^2 \leq 1 \bigr\}
\end{equation}
because of, say, a potential $V$ that goes to $+\infty$ as $\canq$
approaches the cylinder from outside. The effective configuration
space $\Q =\RRR^3 \setminus \cylinder$ has the same fundamental group
$\ZZZ$ as the circle since it is diffeomorphic to $S^1 \times \RRR^+
\times \RRR$ (the fundamental group of a Cartesian product is the
direct product of the fundamental groups, and the half plane $\RRR^+
\times \RRR$ is simply connected). The magnetic field $\magnetic$,
which vanishes outside $\cylinder$ but not inside, is included in the
equations by means of a vector potential $\vA$ with $\nabla \times \vA
= \magnetic$:
\begin{subequations} \label{ie:veA}
\begin{align}
  \sdud{Q_t}{t} &= v^\psi(Q_t) = \hb\, \im \frac{\inpr{\psi}{(\nabla -
      \tfrac{ie}{\hbar} \vA) \psi}}{\inpr{\psi}{\psi}} (Q_t) \\
  i \hb \dud{\psi_t}{t} &= - \tfrac{\hb^2}{2} (\nabla -
  \tfrac{ie}{\hbar} \vA)^2 \psi_t + V \psi_t \label{ie:SeA}
\end{align}
\end{subequations}
with $e$ the charge of the particle.  These equations are in fact best
regarded as instances of the immediate dynamics \eqref{ie:ve} on a
\herm\ bundle for a nontrivial connection on the trivial vector bundle
$\Q \times \CCC$, a \z{point of} view that we will discuss in \Sect{\ref{s:ab3}}.

For now we observe that the vector potential can be gauged away---not
on $\Q$ but on the covering space $\covspa$ (which is diffeomorphic to
$\RRR \times \RRR^+ \times \RRR$). More precisely, the dynamics
\z{is unaffected by}  (i) lift to the
covering space with $\gamma=1$, and (ii) change of gauge. 

Concerning (i), note that $v^\psi$ is the projection of the vector
field
\begin{equation}
  \hat{v}^{\hat{\psi}} = \hb\, \im \frac{\inpr{\hat\psi}
  {(\nabla - \tfrac{ie}{\hbar} \wh\vA) \hat\psi}}
  {\inpr{\hat\psi}{\hat\psi}} 
\end{equation}
where $\hat\psi$ is the lift of $\psi$ (and thus a periodic wave
function with $\gamma = 1$) and evolves according to the lift of
\eqref{ie:SeA},
\begin{equation}
  i \hb \dud{\hat\psi_t}{t} = - \tfrac{\hb^2}{2} (\nabla -
  \tfrac{ie}{\hbar} \wh\vA)^2 \hat\psi_t + \wh{V} \hat\psi_t \,.
\end{equation}
We will now write again $\psi$, rather than $\hat\psi$, for the wave
function on $\covspa$.

Concerning (ii), a change of gauge means the simultaneous replacement
\begin{equation}\label{changegauge}
  \wh\vA(\covq) \to \wh\vA'(\covq) = \wh\vA(\covq) + \nabla f(\covq) \,,
  \quad \psi(\covq) \to \psi'(\covq) =
  e^{ief(\covq)/\hbar} \psi(\covq)
\end{equation}
for an arbitrary function $f: \covspa \to \RRR$. This does not change
the dynamics.  A vector field with vanishing curl, such as $\vA$ on
$\Q$ or $\wh\vA$ on $\covspa$, is a gradient in every simply connected
region; thus, while $\vA$ is locally but not globally a gradient,
$\wh\vA$ is globally a gradient,
\begin{equation}
  \wh\vA = \nabla g\,.
\end{equation}
The (not projectable) function $g: \covspa \to \RRR$ is given \z{by}  
\begin{equation}\label{gintA}
  g(\covq) = \int_{\covq_0}^{\covq} \wh\vA(\covr) \cdot d\covr + C
\end{equation}
\z{for arbitrary $\covq_0 \in \covspa$,} where the integration path is an
arbitrary curve from $\covq_0$ to $\covq$ and $C$ \z{is} a constant
depending on $\covq_0$.  By setting $f= - g$, we can change the gauge in
such a way that $\wh\vA'$ vanishes.

However, the change of gauge affects the periodicity of the wave
function $\psi$: instead of $\psi(\sigma \covq) = \psi (\covq)$ we
\z{have that} 
\begin{equation}\label{e:perconAB2}
  \psi'(\sigma \covq) = \gamma_\sigma \, \psi'(\covq)
\end{equation}
with $\gamma_\sigma = \gamma^{k(\sigma)}$, where $\gamma =
\exp(-ie\Phi/\hbar)$ \z{as}  in \eqref{ABphase}, \z{with} $\Phi$  \z{the}  magnetic
flux given by \eqref{Phidef}, and $k(\sigma) \in \ZZZ$ is the number
of full counterclockwise rotations that the covering transformation
$\sigma$ \z{induces}  on $\covspa$.

To see this, \z{note first that}  $\psi'(\sigma \covq) = \exp(-ieg(\sigma
\covq)/\hbar) \, \psi(\sigma \covq) = \exp(-ieg(\sigma \covq)/\hbar)
\, \psi(\covq)$. Since, by \eqref{gintA}, 
\[
  g(\sigma \covq) =
  \int_{\covq_0}^{\covq} \wh\vA(\covq) \cdot d\covr +
  \int_{\covq}^{\sigma \covq} \wh\vA(\covr) \cdot d\covr + C =
  \int_{\covq}^{\sigma \covq} \wh\vA(\covr) \cdot d\covr + g(\covq)
\]
for arbitrary integration paths with the indicated end points, we have
that 
\begin{equation}
  \psi'(\sigma \covq) = \exp \Bigl( -i \tfrac{e} {\hbar}
  \int_{\hat\closedcurve} \wh\vA(\covr) \cdot d\covr \Bigr) \,
  \psi'(\covq) =: \gamma_\sigma \, \psi'(\covq)
\end{equation}
for arbitrary path $\hat\closedcurve$ from $\covq$ to $\sigma \covq$.
To evaluate the integral and show that it is independent of $\covq$,
note that it agrees with the corresponding integral along the
projected path $\closedcurve = \proj(\hat\closedcurve)$, a loop in
$\Q$:
\begin{equation}
  \int_{\hat\closedcurve} \wh\vA(\covr) \cdot d\covr
  = \int_{\closedcurve} \vA(r) \cdot dr\,,
\end{equation}
which depends only on the homotopy class of $\closedcurve$.  Now
consider for $\closedcurve$ a loop in $\Q$ that surrounds the cylinder
once counterclockwise.  By the Stokes integral formula, the last
integral then agrees with the integral of $\magnetic$ over any surface
$D$ in $\RRR^3$ bounded by $\closedcurve$,
\begin{equation}
  \int_{\hat\closedcurve} \wh\vA(\covr) \cdot d\covr =
  \int_D \magnetic \cdot \boldsymbol{n} \, dA = \Phi \,.
\end{equation}
This completes the proof.

\z{The dynamics can thus}  be described without a vector potential by a
wave function on $\covspa$ satisfying the periodicity condition
\eqref{e:perconAB2}. Ignoring the radial and $q_3$ coordinates,
keeping only the circle $S^1$, yields the model of \Sect{\ref{s:ab}}.

%

\subsection{The Bundle View}
\label{s:ab3}

We re-express our discussion in \Sect{\ref{s:ab2}} of the
Aharonov--Bohm effect in terms of \herm\ bundles.

The dynamics that fundamentally takes place in $\RRR^3$ is the
immediate dynamics, given by \eqref{ie:ve}, for the \herm\ bundle
$E^0$ consisting of the trivial vector bundle $\RRR^3 \times\CCC$, the
local inner product $\inpr{\phi(q)} {\psi(q)}_q = \overline{\phi(q)}
\psi(q)$, and the nontrivial connection whose gradient operator is
$\nabla = \nabla_\mathrm{trivial} - \tfrac{ie}{\hbar} \vA$. (The inner
product is parallel iff $\im \vA=0$.) The curvature of this connection
is proportional to the magnetic field $\magnetic$; \z{therefore}  $E^0$ is
curved but its restriction $E=E^0|_{\Q}$ to the effective
configuration space $\Q = \RRR^3 \setminus \cylinder$ is flat.

Let, for the moment, $E$ be any flat \herm\ line bundle 
over any Riemannian manifold $\Q$.  Its lift $\wh{E}$ is a trivial
\herm\ bundle, like every flat bundle over a simply connected base
manifold.  We obtain the same dynamics (as from $E$) from periodic
sections of $\wh{E}$ with $\gamma =1$.  However, the description of
$\wh{E}$ in which the periodicity condition has trivial topological
factor $\gamma=1$, namely the description as the lift of $E$, is not
the description in which the triviality of $\wh{E}$ is manifest,
namely the description relative to a trivialization, which may change
the topological factor in the periodicity condition as follows.

A trivialization corresponds to a parallel choice of orthonormal basis
in every fiber $\wh{E}_{\covq}$ (i.e., of an identification of
$\wh{E}_{\covq}$ with $\CCC$), and thus to a parallel section $\phi$
of $\wh{E}$ with $\inpr{\phi_{\covq}} {\phi_{\covq}}_{\covq} =1$.
Relative to this trivialization, a section $\psi$ of $\wh{E}$
corresponds to a function $\psi': \covspa \to \CCC$ according to
\begin{equation}\label{psicoophi}
  \psi(\covr) = \psi'(\covr) \,\phi(\covr) \,.
\end{equation}
If $\psi(\sigma \covq) = \psi(\covq)$ corresponding to $\gamma=1$ then
$\psi'$ satisfies the periodicity condition
\begin{equation}
  \psi'(\sigma \covq)  = \holonomy_{\closedcurve}^{-1} \, \psi'(\covq)\,,
\end{equation}
where $\closedcurve$ is any loop in $\Q$ based at $\proj(\covq)$ whose
lift $\hat\closedcurve$ starting at $\covq$ leads to $\sigma \covq$,
and $\holonomy_{\closedcurve}$ is the associated holonomy (which in
this case, with $\mathrm{rank} \,E = 1$, is a complex number of
modulus 1).  This follows from \eqref{psicoophi} by parallel transport
along $\hat\closedcurve$, using that, by parallelity,
$\phi(\sigma\covq) = \holonomy_{\closedcurve} \, \phi(\covq)$.

As a consequence, every dynamics from $\class_0(\Q,E,V)$ for a flat
\herm\ line bundle $E$ exists also in $\class_1(\Q,V)$. In other
words, we can avoid the use of a nontrivial flat \herm\ line bundle if
we use a dynamics of class $\class_1 \setminus \class_0$ with suitable
topological phase factor.

Let us return to the concrete bundle defined in the beginning of this
section. It remains to determine the holonomy.  For a loop
$\closedcurve$ in $\Q$ that surrounds the cylinder once
counterclockwise,
\begin{equation}
  \holonomy_\closedcurve = \exp \Bigl( \tfrac{ie}{\hbar}
  \int_\closedcurve \vA(r) \cdot dr \Bigr)\,.
\end{equation}
Since the integral equals, according to our computation in
\Sect{\ref{s:ab2}}, the magnetic flux $\Phi$, the topological phase
factor is given by $\gamma = \exp(-ie\Phi/\hbar)$.

\section{Vector-Valued Periodic Wave Functions on the
   Covering Space}
\label{s:periodic2}

When the wave function is not a scalar but rather a mapping to a vector
space $\sspa$ of dimension \z{greater than 1}, such as for a particle with
spin, the topological factors can be matrices, forming a unitary
representation of $\fund{\Q}$, as we shall derive presently. The more
complicated case in which $\psi_t$ is a section of a vector bundle is
discussed in \Sect{\ref{s:bundlevalued}}. The possibility of topological
factors given by representations more general than characters was first
mentioned in \cite{Sch81}, Notes to Section 23.3.

\subsection{Vector Spaces}\label{s:vectorvalued}

\z{Suppose that}  the wave functions assume values in a Hermitian vector
space $\sspa$. \z{Then in} a periodicity condition analogous to
\eqref{e:percon},
\begin{equation}\label{e:percon05}
  \psi(\deckt \covq) = \Gamma_\deckt \,\psi(\covq),
\end{equation}
\z{we can}  allow the topological factor $\Gamma_\deckt$ to be
an endomorphism $\sspa \to \sspa$, rather than just a complex number.
By the same argument as in the scalar case, using that $\psi(\covq)$
can be any element of $\sspa$, $\Gamma$ must be a representation of
$\deckg$ on $\sspa$.

It follows from \eqref{e:percon05} that $\nabla \psi(\deckt \covq) =
(\deckt^* \otimes \Gamma_\deckt) \nabla \psi(\covq)$, where $\nabla
\psi(\covq)$ is viewed as an element of $\CCC T_{\covq}\covspa \otimes
\sspa$.  Assume now that, in addition, $\Gamma$ is a \emph{unitary}
representation of $\deckg$. Then the velocity field $\hat{v}^\psi$ on
\covspa\ associated with $\psi$ according to
\begin{equation}
  \hat{v}^\psi (\covq) := \hbar \, \im \,
  \frac{(\psi,\nabla\psi)}{(\psi,\psi)} (\covq)
\end{equation}
is projectable, $\hat{v}^\psi (\deckt \covq) = \deckt^* \hat{v}^\psi
(\covq)$, and gives rise to a velocity field $v^\psi$ on \gencon.  (While
we used unitarity in the scalar case of \Sect{\ref{sec:scalarperiodic}}
only for obtaining the equivariant probability density, we use it here
already for having a projectable velocity field. For this purpose, we could
\z{have allowed} $\Gamma_\deckt$ to be, rather than unitary, a complex
multiple of a unitary \z{endomorphism; unitarity would then be required} in
order to obtain an equivariant density.)

The potential $V$ can \z{now} assume values in the Hermitian endomorphisms of
$\sspa$; the space of endomorphisms can be written $\sspa \otimes
\sspa^*$, so that $V$ is a function $\gencon \to \sspa \otimes
\sspa^*$. We let $\psi$ evolve according to the Schr\"odinger equation
on \covspa,
\begin{equation}\label{e:sch05}
  i\hbar \frac{\partial \psi}{\partial t}(\covq) = -\tfrac{\hbar^2}{2}
  \gD \psi(\covq) + \widehat{V}(\covq) \, \psi(\covq). 
\end{equation}
\z{The periodicity condition}  \eqref{e:percon05} \z{is} preserved
by the evolution \eqref{e:sch05} \z{when}  and only when every
$\Gamma_\deckt$ commutes with every $V(q)$,
\begin{equation}\label{e:commute05}
  \Gamma_\deckt V(q) = V(q) \Gamma_\deckt
\end{equation}
for all $\deckt \in \deckg$ and all $q \in \gencon$.  To see this, note
that $\psi \circ \deckt$ is a solution of \eqref{e:sch05} if $\psi$
is. \z{Thus} \eqref{e:percon05} is preserved if and only if $\Gamma_\deckt
\psi$ satisfies \eqref{e:sch05}, which is the case \z{precisely} when
multiplication by $\Gamma_\deckt$ commutes with the Hamiltonian. Since it
trivially commutes with the Laplacian, the relevant condition is that
$\Gamma_\deckt$ \z{commute}  with the potential $\widehat{V}(\covq)$ at every
$\covq \in \covspa$, or, what amounts to the same, with $V(q)$ at every $q
\in \gencon$.

Given \eqref{e:commute05}, we can let the configuration $Q_t$ move
according to $v^{\psi_t}$,
\begin{equation}\label{e:bohm05}
  \frac{dQ_t}{dt} = v^{\psi_t}(Q_t) = \hbar \, \proj^* \Bigl( \im \,
  \frac{(\psi,\nabla \psi)}{(\psi,\psi)} \Bigr)(Q_t). 
\end{equation}

\z{Since}  $\Gamma$ is a \emph{unitary} representation of
$\deckg$, the motion \eqref{e:bohm05} has an equivariant probability
distribution, namely
\begin{equation}\label{e:equi05}
  \rho(q) = (\psi(\covq),\psi(\covq)). 
\end{equation}
The right hand side does not depend on the choice of $\covq \in
\proj^{-1}(q)$ since, by \eqref{e:percon05}, $(\psi(\deckt
\covq),\psi(\deckt \covq)) = (\Gamma_\deckt \psi(\covq), \Gamma_\deckt
\psi(\covq)) = (\psi(\covq), \psi(\covq))$. Equivariance can be
\z{established} in the same way as in the scalar case.

We define the Hilbert space $L^2(\covspa,\sspa,\Gamma)$ to be the set
of measurable functions $\psi:\covspa \to \sspa$ (modulo changes on
null sets) satisfying \eqref{e:percon05} with
\begin{equation}
  \int_\gencon dq \, (\psi(\covq),\psi(\covq)) < \infty,
\end{equation}
endowed with the scalar product
\begin{equation}
  \langle \phi, \psi \rangle = \int_\gencon dq \,
  (\phi(\covq),\psi(\covq)). 
\end{equation}
Again, the value of the integrand at $q$ is independent of the choice
of $\covq \in \proj^{-1}(q)$.

We summarize the results of our reasoning. 

\begin{assertion}\label{a:vector}
  Given a Riemannian manifold \gencon, a Hermitian vector space
  $\sspa$, and a Hermitian function $V: \gencon \to \sspa
  \otimes \sspa^*$, there is a Bohmian dynamics for each unitary
  representation $\Gamma$ of $\deckg$ on $\sspa$ that commutes
  with all \z{the endomorphisms}  $V(q)$; it is defined by \eqref{e:percon05},
  \eqref{e:sch05}, and \eqref{e:bohm05}, where the wave function
  $\psi_t$ lies in $L^2(\gencon,\sspa,\Gamma)$ and has norm 1. 
\end{assertion}

We define $\class_2 (\gencon, \sspa, V)$ to be the class of Bohmian
dynamics provided by Assertion~\ref{a:vector}. 

The characters $\gamma$ of $\deckg$ (which are in a canonical
one-to-one correspondence with the characters of $\fund{\Q}$) \z{ are} 
contained in Assertion~\ref{a:vector} as special cases of unitary
representations $\Gamma$ \z{by setting}  
\begin{equation}\label{Gammagamma}
  \Gamma_\deckt = \gamma_\deckt \, \id_\sspa \,. 
\end{equation}
These are precisely those \z{unitary} representations $\Gamma$ for which all
$\Gamma_\deckt$ are multiples of the identity.  We define $\class_1
(\gencon, \sspa, V)$ to be the class of those Bohmian dynamics from
$\class_2 (\gencon, \sspa, V)$ arising from a \z{unitary} representation $\Gamma$
of the form \eqref{Gammagamma}, i.e., arising from a character.  This
class contains as many elements as there are characters of
$\fund{\gencon}$, since different characters define different
dynamics; we give a proof of this fact in \cite[Sect.~6.4]{topid1B} \z{(see
also footnote \ref{fn:path})}.
The definition of $\class_1 (\gencon, \sspa, V)$ agrees with that of
$\class_1(\Q,V)$ given in \Sect{\ref{sec:scalarperiodic}} in the sense
that the latter is the special \z{case}   $\sspa = \CCC$,
$\class_1(\Q,V) = \class_1(\Q,\CCC,V)$. Trivially, $\class_0(\Q,
\sspa, V) \subseteq \class_1(\Q, \sspa, V) \subseteq \class_2(\Q,
\sspa, V)$.

\subsection{Remarks}

\begin{enumerate}\addtocounter{enumi}{\theremi}
  
\item The condition that $\Gamma$ \z{be a representation of $\deckg$ that}
commutes with $V$ can alternatively be expressed by saying that $\Gamma$ is
a \z{homomorphism $\deckg\to C(V)$} where $C(V)$ denotes the
\emph{centralizer} of $V$, i.e., the subgroup of $U(\sspa)$ (the unitary
group of $\sspa$) containing all elements that commute with each $V(q)$.
  
\item\label{rem:equiWVGamma} The dynamics defined by $\sspa$, $V$, and
  $\Gamma$ is the same as the one defined by $\sspa'$, $V'$, and
  $\Gamma'$ (another vector space, a potential on $\sspa'$, and a
  representation on $\sspa'$) if there is a unitary isomorphism $U:
  \sspa \to \sspa'$ such that
  \begin{equation}\label{VUVU}
    V' = U V U^{-1}
  \end{equation}
  and
  \begin{equation}\label{GUGU}
    \Gamma' = U \Gamma U^{-1}\,.
  \end{equation}
  
  To see this, define a mapping $\psi \mapsto \psi'$, from
  $L^2(\covspa, \sspa, \Gamma)$ to $L^2(\covspa, \sspa', \Gamma')$, by
  $\psi'(\covq) := U \psi(\covq)$.  Here we use that
  \[
    \psi'(\sigma \covq) = U
    \psi(\sigma \covq) = U \Gamma_\sigma \psi(\covq) =
    U\Gamma_\sigma
    U^{-1} \psi'(\covq) = \Gamma'_\sigma \psi'(\covq)\,.  
  \]
  Since $(-\tfrac{\hbar^2}{2} \gD + \wh{V}') \psi' =
  U(-\tfrac{\hbar^2}{2} \gD + \wh{V})\psi$, $U$ intertwines the time
  evolutions on $L^2(\covspa, \sspa, \Gamma)$ and $L^2(\covspa,
  \sspa', \Gamma')$ based on $V$ and $V'$, i.e., $(\psi')_t =
  (\psi_t)'$.  Since, moreover, at any fixed time $\psi'$ and $\psi$
  lead to the same probability distribution $\rho$ on $\gencon$ and to
  the same velocity fields $\hat{v}^{\psi'} = \hat{v}^{\psi}$ and
  $v^{\psi'} = v^{\psi}$, $\psi'$ and $\psi$ lead to the same
  trajectories with the same probabilities.  That is, the dynamics
  defined by $\sspa,V,\Gamma$ and \z{the one} defined by $\sspa',V',\Gamma'$
  are \z{the same}.
  
\item \label{rmk:equiv} As a consequence of the previous remark, we can use, \z{in Assertion
\ref{a:vector},} representations of the \emph{fundamental group}
$\fund{\gencon}$ instead of representations of the \emph{covering group}
$\deckg$.

  With a unitary representation $\tilde{\Gamma}$ of $\fund{\Q,q}$ on the
  vector space $\sspa$ (for any $q$) there are naturally associated several
  unitary representations $\Gamma(\covq)$ of $\deckg$ on $\sspa$, one for
  each $\covq \in \proj^{-1}(q)$, defined by $\Gamma(\covq) =
  \tilde{\Gamma} \circ \varphi_{\covq}$, \z{i.e., by
  $\Gamma_\tau(\covq)=\tilde{\Gamma}_{\varphi_{\covq}(\tau)}$,} using the
  isomorphism $\varphi_{\covq} : \deckg \to \fund{\gencon,q}$ introduced in
  \Sect{\ref{sec:notation}}. However, these representations $\Gamma(\covq)$
  lead to the same dynamics. To see this, consider two points $\covq, \covr
  \in \proj^{-1}(q)$ \z{with $\covr=\sigma\covq, \ \sigma\in\deckg$. Then
  $\Gamma_{\tau}(\covr) =
  \tilde{\Gamma}_{\varphi_{\covr}(\tau)}=\tilde{\Gamma}_{\varphi_{\covq}(\sigma^{-1}\tau\sigma)}=
  \Gamma_{\sigma^{-1} \tau \sigma}(\covq) = \Gamma_{\sigma}(\covq)^{-1} \,
  \Gamma_{\tau}(\covq) \, \Gamma_{\sigma}(\covq) = U \Gamma_\tau(\covq)
  U^{-1}$ with $U= \Gamma_{\sigma}(\covq)^{-1}$} a unitary endomorphism of
  $\sspa$.  Since $\tilde{\Gamma}$ commutes with $V$ so does $U$, and by
  virtue of \z{Remark~\ref{rem:equiWVGamma}}, $\sspa,V,\Gamma(\covq)$
  \z{defines} the same dynamics as \z{does} $\sspa,V,\Gamma(\covr)$.
  
\item \label{rmk:equivv} As a further consequence of
Remark~\ref{rem:equiWVGamma}, corresponding to the case in which $\sspa' =
\sspa$ and $V' =V$, \z{if $\Gamma' = U\Gamma U^{-1}$ for $U \in C(V)$ (so
that $UVU^{-1} = V$)} then $\sspa$, $V$, and $\Gamma'$ define the same
dynamics as $\sspa$, $V$, and $\Gamma$. Therefore, $\class_2 (\Q, \sspa,
V)$ contains at most as many elements as there are homomorphisms
$\tilde{\Gamma} : \fund{\Q} \to C(V)$ modulo conjugation by elements $U$ of
$C(V)$.
  
\item \label{rmk:ch} The characters\z{---more precisely, the representations of
the form \eqref{Gammagamma}---}commute with all endomorphisms of $\sspa$, and
are thus compatible with \emph{every} potential. All other unitary
representations $\Gamma$ are compatible only with \emph{some}
potentials. \z{(If $\Gamma_\deckt$ is not a multiple of the identity, then
there is a Hermitian endomorphism, which could occur as a $V(q)$ for some
$q$, that does not commute with it.)}

\item \label{rmk:gen} \z{Moreover, characters are the only representations that commute with a potential $V$ when (and, if $\fund{\Q}$
has a nontrivial character, only when) the algebra $\mathrm{Alg}(V(\Q))$
generated by the $V(q)$ is the full endomorphism algebra
$\mathrm{End}(\sspa)$ of $\sspa$, a condition that is satisfied for a
generic potential. Thus, for a generic potential $V$, $\class_2 (\Q, \sspa,
V)=\class_1 (\Q, \sspa, V)$.}

\item \label{rmk:gen2}\z{Similarly, characters are the only representations
that commute with several potentials $V_1, \ldots, V_m$ when the algebra
$\mathrm{Alg} \bigl( V_1(\Q) \cup \ldots \cup V_m(\Q) \bigr)$ generated by
the $V_1(q), \ldots, V_m(q)$ is $\mathrm{End}(\sspa)$.}

\item \label{rmk:dec} \z{Even for $V$ such that
$\mathrm{Alg}(V(\Q))\neq\mathrm{End}(\sspa)$, generically only characters are
necessary. This is because for a generic such $V$ there will be a $q\in\Q$
such that $V(q)$ is nondegenerate. Since $\Gamma_\sigma$ must commute with
$V(q)$, $\Gamma_\sigma$ and $V(q)$ must be simultaneously diagonalizable,
and if $V(q)$ is nondegenerate we have that the representation $\Gamma$ is
diagonal, with diagonal entries $\gamma^{(i)}$ given by characters, in the
basis $|i\rangle\in\sspa$ of eigenvectors of $V(q)$. In other words, the
representation is of the form}

\begin{equation}\label{eq:d} 
\Gamma_\sigma = \sum_i \gamma_{\sigma}^{(i)}P_{\sspa^{(i)}}, 
\end{equation}
\z{where $P_{\sspa^{(i)}}$ is the projection onto the $i$-th eigenspace
$\sspa^{(i)}=\CCC|i\rangle$ of $V(q)$. Moreover, when the $\gamma^{(i)}$'s
are all different, we then have that every $V(r)$ is diagonal in the basis
$|i\rangle$ and the corresponding Schr\"odinger dynamics, of class
$\class_2 (\Q,\sspa,V)$, can be decomposed into a direct sum of dynamics of
class $\class_1(\Q,\sspa^{(i)},V^{(i)})$, given by characters, where
$V^{(i)}$ is the action of $V$ on $\sspa^{(i)}$. Thus, the set of dynamics
corresponding to representations $\Gamma$ of the form \eqref{eq:d} could be
denoted}

\begin{equation}
\bigoplus_i\class_1(\Q,\sspa^{(i)},V^{(i)}).
\end{equation} 
\z{When the $\gamma^{(i)}$'s are not distinct, a similar decomposition
holds, with the sum over $i$ replaced by the sum over the distinct
characters $\gamma$ and with the $\sspa^{(i)}$'s replaced by the spans of
the $\sspa^{(i)}$'s corresponding to the same $\gamma$. }

\item \label{rmk:irred} \z{In the situation described in the previous
remark, the representation $\Gamma$ on $\sspa$ is reducible, as it clearly
is when it is given by a character (unless $\dim\sspa=1$). In fact, by  Schur's lemma,  $\Gamma$
can be irreducible only when the potential $V$ is a scalar, i.e., of the
form $V(q)=\tilde{V}(q)\mathrm{Id}_{\sspa}$ with $\tilde{V}(q)\in
\RRR$.}

\item \label{rmk:V*}\z{We have so far considered the possible Bohmian
dynamics associated with a configuration space $\Q$, a Hermitian vector
space $\sspa$ and a Hermitian function $V: \gencon \to \sspa \otimes
\sspa^*$, and have argued that we have one such dynamics for each
representation $\Gamma$ of $\deckg$ that commutes with $V$. Let us now
consider the class $\class_2 (\Q,\sspa,\Gamma)$ of possible Bohmian
dynamics associated with a configuration space $\Q$, a Hermitian vector
space $\sspa$, and a representation $\Gamma$ of $\deckg$. There is of
course one such dynamics for every choice of $V$ that commutes with
$\Gamma$ but there are more. In fact, in addition to these there is also
a dynamics for every Hermitian function $V^*: \covspa \to \sspa \otimes
\sspa^*$ satisfying}

\begin{equation}\label{Vcov}
V^*(\sigma\covq)=\Gamma_\sigma
V^*(\covq)\Gamma_{\sigma}^{-1},
\end{equation}

\z{a dynamics involving a potential $V^*(\covq)$ on $\covspa$ that need not
be the lift of any potential $V$ on $\Q$. If $\Gamma$ is given by a
character then $V^*$ must in fact be the lift of a potential $V$ on $\Q$,
$V^*(\covq)=V(\proj(q))$, and we obtain nothing new, but when $\Gamma$ is
not given by a character, many new possibilities occur.}

\setcounter{remi}{\theenumi}
\end{enumerate}

\subsection{Examples}\label{s:ex}

Let us give an example of matrices as topological factors: a
higher-dimensional version of the Aharonov--Bohm effect. We may
replace the vector potential in the Aharonov--Bohm setting by a
non-abelian gauge field (\`a la Yang--Mills) whose field strength
(curvature) vanishes outside the
cylinder $\cylinder$ but not inside; the value space $\sspa$ (now
corresponding not to spin but to, say, quark color) has dimension
greater than one, and the difference between two wave packets that
have passed the cylinder $\cylinder$ on different sides is in general,
rather than a phase, a unitary endomorphism $\Gamma$ of $\sspa$. 

A more practical version is provided by the Aharonov--Casher variant
\cite{AC84} of the Aharonov--Bohm effect, according to which a neutral spin-1/2
particle that carries a magnetic moment $\mu$ acquires a
nontrivial phase while encircling a charged wire $\cylinder$. Start with the Dirac
equation for a neutral particle with nonzero magnetic moment $\mu$
(such as a neutron),
\begin{equation}\label{Dirac}
  i\hbar \gamma^\mu \partial_\mu \psi = m\psi + \tfrac{1}{2} \mu
  F^{\mu\nu} \sigma_{\mu\nu} \psi \,,
\end{equation}
where $\psi: \RRR^4 \to \CCC^4$, $\gamma^\mu$ are the four Dirac
matrices, $F^{\mu\nu}$ is the field tensor of the external
electromagnetic field, and $\sigma_{\mu\nu} = \gamma_\mu \gamma_\nu -
\gamma_\nu \gamma_\mu$. The last term in \eqref{Dirac} should be
regarded as  phenomenological.  Consider now the nonrelativistic
limit, in which the wave function assumes values in spin space $\sspa
= \CCC^2$, acted upon by the vector $\boldsymbol{\sigma}$ of spin
matrices.  Suppose that the magnetic field is zero and the electric
field $\electric$ is generated by a charge distribution $\varrho(\canq)$ inside
$\cylinder$ which is 
invariant under translations in the direction $\boldsymbol{e}\in
\RRR^3$, $\boldsymbol{e}^2=1$ in which the wire is oriented. Then
the charge per unit length $\lambda$ is given by the integral
\begin{equation} \lambda = \int_D \varrho(\canq)\, dA
\end{equation} over the cross-section disk $D$ in any plane
perpendicular to $\boldsymbol{e}$. The Hamiltonian is \cite{AC84}
\begin{equation}\label{ACH}
  H = -\tfrac{\hbar^2}{2m} \bigl( \nabla - \tfrac{i\mu}{\hbar}
  \electric \times \boldsymbol{\sigma} \bigr)^2 -
  \tfrac{\mu^2}{m} \electric^2 \,,
\end{equation}
where $\times$ denotes the vector product in $\RRR^3$. This looks like
a Hamiltonian $\rawH$ based on a nontrivial connection $\nabla =
\nabla_\mathrm{trivial} - \tfrac{i\mu}{\hbar} \electric \times
\boldsymbol{\sigma}$ on the vector bundle $\RRR^3 \times \CCC^2$.  
The restriction of this connection,
outside of $\cylinder $, to any plane $\Sigma$ orthogonal to the wire turns
out to be flat\footnote{The curvature is $\Omega = d_\mathrm{trivial}
\boldsymbol{\omega} + \boldsymbol{\omega} \wedge \boldsymbol{\omega} $ with
$\boldsymbol{\omega} = -i\frac{\mu}{\hbar} \electric \times
\boldsymbol{\sigma}$. The 2-form $\Omega$ is dual to the vector
$\nabla_\mathrm{trivial} \times\boldsymbol{\omega} + \boldsymbol{\omega}
\times \boldsymbol{\omega} =i\frac{\mu}{\hbar}
(\nabla\cdot\electric)\boldsymbol{\sigma} -
i\frac{\mu}{\hbar}(\boldsymbol{\sigma}\cdot\nabla)\electric - 2i
(\frac{\mu}{\hbar})^2(\boldsymbol{\sigma} \cdot \electric) \electric.$
Outside the wire, the first term vanishes and, noting that
$\electric\cdot\boldsymbol{e} =0,$ the other two terms have vanishing
component in the direction of $\boldsymbol{e}$ and thus vanish when
integrated over any region within an orthogonal plane.} so that its
restriction to the intersection $\Q$ of $\RRR^3\setminus\cylinder$ with the
orthogonal plane can be replaced, as in the Aharonov--Bohm case, by the
trivial connection if we introduce a periodicity condition on the wave
function with the topological factor
\begin{equation} \Gamma_1 = \exp \Bigl(-\frac{4\pi i\mu\lambda}{\hbar}\,
\boldsymbol{e}\cdot\boldsymbol{\sigma} \Bigr) \,.  \end{equation}
In this way we obtain a representation $\Gamma : \fund{\Q}
\to SU(2)$ that is not given by a character.
(For further discussion of the link between gauge connections and
topological factors, see \cite{topid1B}.)

Though the $\Gamma$'s are matrices in the above examples, the
representation is still abelian since $\fund{\Q} \cong \ZZZ$ is an
abelian group. To obtain a non-abelian representation, let $\Q$ be
$\RRR^3$ minus two disjoint solid cylinders; its fundamental group is
isomorphic to the non-abelian group $\ZZZ \ast \ZZZ$ where $\ast$
denotes the free product of groups; it is generated by two loops,
$\sigma_1$ and $\sigma_2$, each surrounding one of the cylinders.
Using again non-abelian gauge fields, one can arrange that the
matrices $\Gamma_{\sigma_i}$ corresponding to $\sigma_i$, $i=1,2$, do
not commute with each other.

Another example, concerning a system of $N$ spin 1/2 fermions: $\Q=\nrd$
(whose fundamental group is the permutation group $S_N$),
$\sspa=\bigotimes_{i=1}^N\CCC^{2}$, and
$\Gamma_\sigma=\mathrm{sgn}(\sigma)R_\sigma$, where $ R_\sigma$ is the natural
action of permutations on the tensor product. The Pauli spin interaction is
well defined on $\covspa$ (but not on $\Q$) for this $\sspa$ (unlike for
the natural spin bundle \eqref{spinbundle}). It is given by
\begin{equation}\label{pauli1} 
V^*(\covq)=-\mu\sum_i \magnetic(\canq_i) \cdot
\boldsymbol{\sigma}_i 
\end{equation}
with $\boldsymbol{\sigma}_i$ the vector of spin matrices acting on the
$i$-th component of the tensor product. $V^*$ is not the lift of any
potential $V$ on $\Q$ (since there is no continuous section of the bundle,
over $\Q=\nrd$, of maps $q\mapsto\{1,\dots,N\}$). This is an example of
class $\class_2(\Q,\sspa,\Gamma)$, see Remark \ref{rmk:V*}.

\subsection{Vector Bundles}
\label{s:bundlevalued}

We now consider wave functions that are sections of vector bundles.
The topological factors will be expressed as \emph{periodicity
  sections}, i.e., parallel unitary sections of the endomorphism
bundle indexed by the covering group \z{and}  satisfying a certain composition
law, or, equivalently, as \emph{twisted representations} of
$\fund{\Q}$.

If $E$ is a vector bundle over \gencon, then the lift of $E$, denoted
by $\wh{E}$, is a vector bundle over \covspa; the fiber space at
\covq\ is defined to be the fiber space of $E$ at $\baseq$,
$\wh{E}_{\covq} \defi E_{\baseq}$, where $\baseq = \proj(\covq)$.  It
is important to realize that with this construction, it makes sense to
ask whether $v \in \wh{E}_{\covq}$ is equal to $w \in \wh{E}_{\covr}$
whenever $\covq$ and $\covr$ are elements of the same covering fiber.
Equivalently, $\wh{E}$ is the pull-back of $E$ through $\proj: \covspa
\to \Q$.  As a particular example, the lift of the tangent bundle of
\gencon\ to \covspa\ is canonically isomorphic to the tangent bundle
of \covspa.  Sections of $E$ or $E\otimes E^*$ can be lifted to
sections of $\wh{E}$ respectively $\wh{E} \otimes \wh{E}^*$.

If $E$ is a \herm\ vector bundle, then so is $\wh{E}$. The wave
function $\psi$ that we consider here is a section of $\wh{E}$, so
that the $\covq$-dependent Hermitian vector space $\wh{E}_{\covq}$
replaces the fixed Hermitian vector space $\sspa$ of the previous
subsection. $V$ is a section of the bundle $E \otimes E^*$, i.e.,
$V(q)$ is an element of $E_q \otimes E_q^*$. To indicate that every
$V(q)$ is a Hermitian endomorphism of $E_q$, we say that $V$ is a
\z{Hermitian section} of $E \otimes E^*$.

Since $\psi(\deckt \covq)$ and $\psi(\covq)$ lie in the same space
$E_q = \wh{E}_{\covq}= \wh{E}_{\deckt \covq}$, a periodicity condition
can be of the form
\begin{equation}\label{e:percon06}
  \psi(\deckt \covq) = \Gamma_\deckt(\covq) \, \psi (\covq)
\end{equation}
for $\deckt \in \deckg$, where $\Gamma_\deckt(\covq)$ is an
endomorphism $E_q \to E_q$.  By the same argument as in
\eqref{e:cccargument}, the condition for \eqref{e:percon06} to be
possible, if $\psi(\covq)$ can be any element of $\wh{E}_{\covq}$, is
the composition law
\begin{equation}\label{e:compo06}
  \Gamma_{\deck1 \deck2}(\covq) = \Gamma_{\deck1} (\deck2 \covq) \,
  \Gamma_{\deck2} (\covq). 
\end{equation}
Note that this law differs from the one $\Gamma(\covq)$ would satisfy
if it were a representation, which reads $\Gamma_{\sigma_1 \sigma_2}
(\covq) = \Gamma_{\sigma_1} (\covq) \, \Gamma_{\sigma_2} (\covq)$, \z{since
in general $\Gamma (\sigma\covq)$ need not be the same as $\Gamma (\covq)$} .

For periodicity \eqref{e:percon06} to be preserved under the
Schr\"odinger evolution,
\begin{equation}\label{e:sch06}
  i\hbar \frac{\partial \psi}{\partial t} (\covq) = -\tfrac {\hbar^2}
  {2} \gD \psi(\covq) + \wh{V}(\covq) \, \psi(\covq),
\end{equation}
we need that multiplication by $\Gamma_\deckt (\covq)$ \z{commute}  with
the Hamiltonian. Observe that
\begin{equation}\label{HGamma}
  [H,\Gamma_\deckt]\psi(\covq) = -\tfrac{\hbar^2}{2} (\gD
  \Gamma_\deckt(\covq)) \psi(\covq) - \hbar^2 (\nabla
  \Gamma_\deckt(\covq)) \cdot (\nabla \psi(\covq)) +
  [\wh{V}(\covq),\Gamma_\deckt(\covq)] \, \psi(\covq). 
\end{equation}
Since we can choose $\psi$ such that, for any one particular $\covq$,
$\psi(\covq)=0$ and $\nabla \psi(\covq)$ is any element of $\CCC T_{\covq}
\covspa \otimes E_q$ we like, we must have that
\begin{equation}\label{e:parallel06}
  \nabla \Gamma_\deckt(\covq) =0
\end{equation}
for all $\deckt\in \deckg$ and all $\covq \in \covspa$, \z{i.e., that
$\Gamma_\sigma$ is parallel.} . Inserting this
in \eqref{HGamma}, the first two terms on the right hand side
vanish. Since we can choose for $\psi(\covq)$ any element of $E_q$ we
like, we must have that
\begin{equation}\label{e:commute06}
  [\wh{V}(\covq),\Gamma_\deckt(\covq)]=0
\end{equation}
for all $\deckt\in \deckg$ and all $\covq \in \covspa$.  Conversely,
assuming \eqref{e:parallel06} and \eqref{e:commute06}, we obtain that
$\Gamma_\deckt$ commutes with $H$ for every $\deckt\in \deckg$, so
that the periodicity \eqref{e:percon06} is preserved.

{}From \eqref{e:percon06} and \eqref{e:parallel06} it follows that
$\nabla \psi(\deckt \covq) = (\deckt^* \otimes \Gamma_\deckt (\covq))
\nabla \psi(\covq)$. If every $\Gamma_\deckt(\covq)$ is
\emph{unitary}, as we assume from now on, the velocity field
$\hat{v}^\psi$ on \covspa\ associated with $\psi$ according to
\begin{equation}
  \hat{v}^\psi (\covq) := \hbar \, \im \,
  \frac{(\psi,\nabla\psi)}{(\psi,\psi)} (\covq)
\end{equation}
is projectable, $\hat{v}^\psi(\deckt \covq) = \deckt^*
\hat{v}^\psi(\covq)$, and gives rise to a velocity field $v^\psi$ on
\gencon. We let the configuration move according to $v^{\psi_t}$,
\begin{equation}\label{e:bohm06}
  \frac{dQ_t}{dt} = v^{\psi_t}(Q_t) = \hbar \, \proj^* \Bigl( \im \,
  \frac{(\psi,\nabla \psi)}{(\psi,\psi)} \Bigr) (Q_t). 
\end{equation}

\begin{defn}
  Let $E$ be a \herm\ bundle over the manifold $\Q$.  A
  \emph{periodicity section} $\Gamma$ over $E$ is a family indexed by
  $\deckg$ of unitary parallel sections $\Gamma_\sigma$ of $\wh{E}
  \otimes \wh{E}^*$ satisfying the composition law \eqref{e:compo06}.
\end{defn}

Since $\Gamma_\deckt(\covq)$ is unitary, one sees as before that the
probability distribution
\begin{equation}\label{e:equi06}
  \rho(q) = (\psi(\covq),\psi(\covq))
\end{equation}
does not depend on the choice of $\covq \in \proj^{-1}(q)$ and is
equivariant. 

As usual, we define for any periodicity section $\Gamma$ the Hilbert
space $L^2(\covspa, \wh{E}, \Gamma)$ to be the set of measurable
sections $\psi$ of $\wh{E}$ (modulo changes on null sets) satisfying
\eqref{e:percon06} with
\begin{equation}
  \int_\gencon dq \, (\psi(\covq),\psi(\covq)) < \infty,
\end{equation}
endowed with the scalar product
\begin{equation}
  \langle \phi, \psi \rangle = \int_\gencon dq \,
  (\phi(\covq),\psi(\covq)). 
\end{equation}
As before, the value of the integrand at $q$ is independent of the
choice of $\covq \in \proj^{-1}(q)$.

We summarize the results of our reasoning. 

\begin{assertion}\label{a:bundle}
  Given a \herm\ bundle $E$ over the Riemannian manifold $\gencon$ and
  a \z{Hermitian section} $V$ of $E \otimes E^*$, there is a
  Bohmian dynamics for each periodicity section $\Gamma$ commuting
  (pointwise) with $\wh{V}$ \z{(cf. (\ref{e:commute06}))}; it is defined by \eqref{e:percon06},
  \eqref{e:sch06}, and \eqref{e:bohm06}, where the wave function
  $\psi_t$ lies in $L^2(\covspa, \wh{E}, \Gamma)$ and has norm 1.
\end{assertion}

The situation of \Sect{\ref{s:vectorvalued}}, where the wave function
assumed values in a fixed Hermitian space $\sspa$ instead of a bundle,
corresponds to the trivial \herm\ bundle $E = \gencon \times \sspa$
(i.e., with the trivial connection, for which parallel transport is
the identity on $\sspa$). Then, parallelity \eqref{e:parallel06}
implies that $\Gamma_\deckt (\covr) = \Gamma_\deckt (\hat{q})$ for any
$\covr, \hat{q} \in \covspa$, or $\Gamma_\deckt (\hat{q}) =
\Gamma_\deckt$, so that \eqref{e:compo06} becomes the usual
composition law $\Gamma_{\deckt_1 \deckt_2} = \Gamma_{\deckt_1}
\Gamma_{\deckt_2}$. As a \z{consequence}, $\Gamma$ is a unitary
 \z{representation of $\deckg$} , and Assertion~\ref{a:vector} is a
special case of Assertion~\ref{a:bundle}.

Every character $\gamma$ of \z{$\deckg$ (or of $\fund{\Q}$)}  defines a periodicity section
by setting 
\begin{equation}\label{Gammagamma2} 
\Gamma_\sigma (\covq) := \gamma_\sigma
\id_{\wh{E}_{\covq}}. 
\end{equation} 
It commutes with every potential $V$.
Conversely, a periodicity section $\Gamma$ that commutes with every
potential must be such that every $\Gamma_\sigma (\covq)$ is a
multiple of the identity, $\Gamma_\sigma (\covq) = \gamma_\sigma
(\covq) \, \id_{\wh{E}_{\covq}}$. By unitarity, $|\gamma_\sigma| =1$;
by parallelity \eqref{e:parallel06}, $\gamma_\sigma (\covq) =
\gamma_\sigma$ must be constant; by the composition law
\eqref{e:compo06}, $\gamma$ must be a homomorphism, and thus a
character.

We define $\class_2 (\gencon, E, V)$ to be the class of Bohmian
dynamics provided by Assertion~\ref{a:bundle}.  We define $\class_1
(\gencon, E, V)$ to be the class of those Bohmian dynamics from
$\class_2 (\gencon, E, V)$ arising from characters: The class
$\class_1 (\Q, E, V)$ contains at most\footnote{For nontrivial \herm\
  bundles, different characters can lead to the same dynamics; we
  \z{gave}  an example in footnote~\ref{ft:charequivalence}.} as many
elements as there are characters of $\fund{\gencon}$.  These
definitions agree with the definitions of $\class_1 (\Q, \sspa, V)$
and $\class_2 (\Q, \sspa, V)$ given in \Sect{\ref{s:vectorvalued}} in
the sense that $\class_1 (\Q, \sspa, V) = \class_1 (\Q, E, V)$ and
$\class_2 (\Q, \sspa, V) = \class_2 (\Q, E, V)$ when $E$ is taken to
be the trivial bundle $\Q \times \sspa$.

\z{We briefly indicate} how a periodicity section $\Gamma$ corresponds to
something like a representation of $\fund{\Q}$, in fact to a
\z{(holonomy-)} twisted representation of $\fund{\Q}$.  Fix a $\covq \in
\covspa$. \z{Then $\deckg$ can be identified with
$\fund{\Q}=\fund{\Q,\proj(\covq)}$ via $\varphi_{\covq}$.}  Since the sections $\Gamma_\sigma$ of $\wh{E} \otimes \wh{E}^*$
are parallel, $\Gamma_\sigma(\covr)$ is determined for every $\covr$ by
$\Gamma_\sigma(\covq)$. \z{(Note in particular that the parallel transport
$\Gamma_\sigma(\tau\covq)$ of $\Gamma_\sigma(\covq)$ from $\covq$ to
$\tau\covq, \tau\in \deckg$, may differ from $\Gamma_\sigma(\covq)$.)} Thus,
the periodicity section $\Gamma$ is completely determined by the
endomorphisms $\Gamma_\sigma := \Gamma_\sigma(\covq)$ of $E_q$, $\sigma \in
\deckg$, which satisfy the composition law
\begin{equation}\label{twistedrep}
  \Gamma_{\sigma_1 \sigma_2} = \holonomy_{\closedcurve_2} \Gamma_{\sigma_1}
  \holonomy_{\closedcurve_2}^{-1} \Gamma_{\sigma_2}\,,
\end{equation}
\z{where $\closedcurve_2$}  is any loop in $\Q$ based at $\proj(\covq)$ whose
lift starting at $\covq$ leads to $\sigma_2 \covq$, and
$\holonomy_{\closedcurve_2}$ is the associated holonomy endomorphism of
$E_q$.  Since \eqref{twistedrep} is not the composition law
$\Gamma_{\sigma_1 \sigma_2} = \Gamma_{\sigma_1} \Gamma_{\sigma_2}$ of
a representation, the $\Gamma_\sigma$ \z{form, not}  a representation of
$\fund{\Q}$, \z{ but}  what we call a \emph{twisted representation}. See
\cite{topid1B} for further discussion.

\subsection{\z{Further Remarks}}
\label{s:exbundlevalued}

\begin{enumerate}\addtocounter{enumi}{\theremi}

\item \label{rem:equiWVGamma2}The dynamics defined by $E$, $V$, and
$\Gamma$ is the same as the one defined by $E'$, $V'$, and $\Gamma'$
(another \herm\ bundle, a potential on $E'$, and a periodicity section over
$E'$) if there is a unitary parallel section $U$ of $\wh{E}' \otimes
\wh{E}^*$ such that
\begin{equation}
  \wh{V}'(\covq) = U(\covq)\, \wh{V}(\covq)\, U(\covq)^{-1}
\end{equation}
and
\begin{equation}
  \Gamma'_\sigma(\covq) = U(\sigma \covq)\, \Gamma_\sigma (\covq)\,
  U(\covq)^{-1}.
\end{equation}
To see this, define a mapping $\psi \mapsto \psi'$, from $L^2(\covspa,
\wh{E}, \Gamma)$ to $L^2(\covspa, \wh{E}', \Gamma')$, by $\psi'(\covq)
:= U(\covq)\, \psi(\covq)$.  Here we use that 
\[
  \psi'(\sigma \covq) =
  U(\sigma \covq)\, \psi(\sigma \covq) = U(\sigma \covq) \,
  \Gamma_\sigma(\covq)\, \psi(\covq) 
\]
\[
= U(\sigma \covq)\,
  \Gamma_\sigma(\covq)\, U(\covq)^{-1} \, \psi'(\covq) =
  \Gamma'_\sigma (\covq)\, \psi'(\covq) \,.
\]
Since, by \z{the}  parallelity of $U$, $(-\tfrac{\hbar^2}{2} \gD + \wh{V}')
\psi' = U(-\tfrac{\hbar^2}{2} \gD + \wh{V})\psi$, $U$ intertwines the
time evolutions on $L^2(\covspa, \wh{E}, \Gamma)$ and $L^2(\covspa,
\wh{E}', \Gamma')$ based on $V$ and $V'$, i.e., $(\psi')_t =
(\psi_t)'$.  Since, moreover, at any fixed time $\psi'$ and $\psi$
lead to the same probability distribution $\rho$ on $\gencon$ (by \z{the} 
unitarity of $U$) and to the same velocity fields $\hat{v}^{\psi'} =
\hat{v}^{\psi}$ and $v^{\psi'} = v^{\psi}$ (by \z{the} parallelity and
unitarity of $U$), $\psi'$ and $\psi$ lead to the same trajectories
with the same probabilities.  That is, the dynamics defined by
$E,V,\Gamma$ and \z{the one}  defined by $E',V',\Gamma'$ are \z{the same}.

\item \z{As a consequence of the previous remark, we can use, in Assertion
\ref{a:bundle}, twisted representations $\tilde{\Gamma}$ of the fundamental group
$\fund{\gencon}$, satisfying}

\begin{equation} 
 \tilde{\Gamma}_{\closedcurve_1\closedcurve_2} =
h_{\closedcurve_2}\tilde{\Gamma}_{\closedcurve_1}h_{\closedcurve_2}^{-1}\tilde{\Gamma}_{\closedcurve_2},
\end{equation}

\z{instead of periodicity sections (twisted representations
of the covering group $\deckg$).} 

\item \z{As a further consequence of Remark~\ref{rem:equiWVGamma2},
corresponding to the case in which $E' = E$ and $V' =V$, if
$\Gamma'_\sigma(\covq) = U(\sigma \covq)\, \Gamma_\sigma (\covq)\,
U(\covq)^{-1}$ for a unitary parallel section $U$ of $\wh{E}' \otimes
\wh{E}^*$ that commutes with $\wh{V}$ then $E$, $V$, and $\Gamma'$ define
the same dynamics as $E$, $V$, and $\Gamma$. Therefore, $\class_2 (\Q, E,
V)$ contains at most as many elements as there are twisted representations
$\tilde{\Gamma}$ of $\fund{\Q}$, i.e., periodicity sections $\Gamma$ over
$E$, that commute with $\wh{V}$, modulo conjugation by such $U$'s.}

\item \label{rmk:chb}\z{As we have already seen, the characters---the
periodicity sections of the form \eqref{Gammagamma2}---are compatible with
every potential, and all other periodicity sections are compatible only
with some potentials.}

\item \label{rmk:genb} \z{A potential $V$ does not commute with any
periodicity section save the characters when (and, if $\fund{\Q}$
has a nontrivial character, only when) for arbitrary $q \in \Q$,  
\begin{equation}\label{gen} 
\mathrm{Alg} \bigl( V(\Q)_q \cup \Theta_q \bigr) =
    \mathrm{End} (E_q) \,,
\end{equation} 
where
$V(\Q)_q = \bigl\{P^{-1}_\opencurve V(r) P_\opencurve : r \in \Q,
    \opencurve \text{ a curve from } q \text{ to }r \bigr\}$ with
    $P_\opencurve: E_q \to E_r$ denoting parallel transport, and $\Theta_q
    = \{\holonomy_\closedcurve : \closedcurve$ a contractible
    loop based at q\} with $\holonomy_\closedcurve = P_\closedcurve$,
    the holonomy of $\closedcurve$. This follows from the fact that
    a periodicity section, by parallelity, must commute with $\Theta_q$.
The condition \eqref{gen} holds, for
example, for the potential occurring in the Pauli equation for $N$
identical particles with spin,}
\begin{equation}\label{pauli2} 
V(q) = -\mu\sum_{\canq \in q} \magnetic(\canq) \cdot
\boldsymbol{\sigma}_{\canq} 
\end{equation} 
on the spin bundle \eqref{spinbundle} over $\nrtre$, with
$\boldsymbol{\sigma}_{\canq}$ the vector of spin matrices acting on the
spin space of the particle at $\canq$, provided merely that the magnetic
field $\magnetic$  is not parallel. Thus, for a generic potential $V$,
$\class_2 (\Q, E, V)=\class_1 (\Q, E, V)$.



\item \z{A periodicity section $\Gamma$ defining a Bohmian dynamics of class
$\class_2 (\Q,E,V)$ can be irreducible only when the potential $V$ is a
scalar. (When $\Gamma$ is reducible, its decomposition may involve
sub-bundles $E^{(i)}$ of  $\wh{E}$ that are not the lifts of any
sub-bundles of $E$.)}

\item \z{Consider the class $\class_2 (\Q,E,\Gamma)$ of possible Bohmian
dynamics associated with a Riemannian manifold $\Q$, a Hermitian bundle $E$
over $\Q$, and a periodicity section $\Gamma$ over  $E$. There is one
such dynamics for every choice of Hermitian section $V^*$ of $\wh{E}
\otimes \wh{E}^*$ satisfying}

\begin{equation}\label{Vcov2}
V^*(\sigma\covq)=\Gamma_{\sigma}(\covq)
V^*(\covq)\Gamma_{\sigma}(\covq)^{-1},
\end{equation}

\z{a dynamics involving a potential $V^*(\covq)$ on $\covspa$ that need not
be the lift of any potential $V$ on $\Q$.}

\item For a \emph{generic} (curved) \herm\ bundle $E$, \z{and any fixed
$V$} we have that $\class_2(\gencon, E,V) = \class_1(\gencon,E,V)$; in
other words, there are no more possibilities than the characters.  This
follows from the fact that, generically, for every unitary endomorphism $U$
of $E_q$ there is a contractible curve $\closedcurve$ based at $q$ whose
holonomy  is $U$.
That is, $\Theta_q$ is the full unitary group of $E_q$,
    and thus, by \eqref{gen},  all periodicity sections
    correspond to characters.

\item \z{Consider a dynamics of class $\class_2(\Q,\sspa,\Gamma)$ or class
$\class_2(\Q,E,\Gamma)$, given by a potential $V^*$ on $\covspa$ satisfying
(\ref{Vcov}), respectively (\ref{Vcov2}). We show in \cite{topid1B} that
there is a Hermitian bundle $E'$ over $\Q$ that is locally isomorphic to
$\Q \times \sspa$, respectively $E$, such that this dynamics, of class
$\class_2$, coincides with the dynamics of class $\class_0(\Q,E',V)$,
i.e., the immediate dynamics for $\Q$, $E'$, and a potential $V$ on
$\Q$. For example, the dynamics associated with $\Q=\nrd$,
$\sspa=\bigotimes_{i=1}^N\CCC^{2s+1}$, and $V^*$ on $\covspa$ given by
(\ref{pauli1}) coincides with the dynamics associated with $\nrtre$, the
natural spin bundle $E'$ \eqref{spinbundle} over $\nrtre$, and the Pauli
interaction $V$ on $\Q$ given by (\ref{pauli2}). (By ``dynamics'' here we
refer to the evolution of the configuration.)}

\setcounter{remi}{\theenumi}
\end{enumerate}

\subsection{Examples Involving Vector Bundles}

We close this section with two examples of topological factors on vector bundles. 

The most important example is provided by identical particles {\it with
spin}. In fact, for this case, Assertion~\ref{a:bundle} entails the same
conclusions we arrived at in Remark \ref{rem:idnospin}, the alternative
between bosons and fermions, even for particles with spin. To understand
how this comes about, consider the potential occurring in the Pauli
equation \eqref{pauli2} for $N$ identical particles with spin, on the spin
bundle \eqref{spinbundle} over $\nrtre$, and observe that the algebra
generated by $\{V(q)\}$ arising from all possible choices of the magnetic
field $\magnetic$ is $\mathrm{End}(E_q)$. Thus the only holonomy-twisted
representations that define a dynamics for all magnetic fields (or even for
a single magnetic field provided only that it is not parallel, see Remark
\ref{rmk:genb}) are those given by a character.

Our last example involves a holonomy-twisted representation $\Gamma$ that
is not a representation in the ordinary sense.  Consider $N$ fermions, each
as in the examples at the beginning of Section \ref{s:ex}, moving in
$M=\RRR^3\setminus \cup_i\cylinder_i$, where $\cylinder_i$ are one or more
disjoint solid cylinders. More generally, consider $N$ fermions, each
having 3-dimensional configuration space $M$ and value space $W$ (which may
incorporate spin or ``color'' or both).  Then the configuration space $\Q$
for the $N$ fermions is the set ${}^N\! M$ of all $N$-element subsets of
$M$, with universal covering space $\covspa=\widehat {{}^N\! M} =
{\widehat{M}}^N\setminus \Delta$ with $\Delta$ the extended diagonal, the
set of points in ${\widehat{M}}^N$ whose projection to $M^N$ lies in its
coincidence set. Every diffeomorphism $\sigma\in Cov(\widehat{{}^N\! M},
{}^N\! M)$ can be expressed as a product
\begin{equation}\label{prod}
\sigma=p\tilde\sigma
\end{equation}
where $p \in S_N$ and $\tilde\sigma =
(\sigma^{(1)},\dots,\sigma^{(N)})\in Cov(\widehat{M},M)^N $ and these act on
$\covq=(\hat\canq_1,\dots, \hat\canq_N)$ $\in\widehat{M}^N$ as follows:
\begin{equation}\label{tildesigmaq}
\tilde\sigma\covq=(\sigma^{(1)}\hat\canq_1,\dots, \sigma^{(N)}\hat\canq_N)
\end{equation}
and
\begin{equation}\label{pq}
p\covq=(\hat\canq_{p^{-1}(1)},\dots, \hat\canq_{p^{-1}(N)}).
\end{equation}
Thus
\begin{equation}\label{sigmaq}
\sigma\covq=(\sigma^{(p^{-1}(1))}\hat\canq_{p^{-1}(1)},\dots, \sigma^{(p^{-1}(N))}\hat\canq_{p^{-1}(N)}).
\end{equation}
Moreover, the representation (\ref{prod}) of $\sigma$ is unique. Thus,
since
\begin{equation}\label{sdp}
\sigma_1\sigma_2=p_1\tilde\sigma_1p_2\tilde\sigma_2=(p_1p_2)(p_2^{-1}\tilde\sigma_1p_2\tilde\sigma_2)
\end{equation}
with
$p_2^{-1}\tilde\sigma_1p_2=(\sigma_1^{(p_2(1))},\dots,\sigma_1^{(p_2(N))})\in Cov(\widehat{M},M)^N$,
we find that
$Cov(\widehat{{}^N\! M}, {}^N\! M)$ is a semidirect product of $S_N$ and
$Cov(\widehat{M},M)^N$, with product
given by
\begin{equation}\label{sprod}
\sigma_1\sigma_2=(p_1,\tilde\sigma_1)(p_2,\tilde\sigma_2)=(p_1p_2,p_2^{-1}\tilde\sigma_1 p_2\tilde\sigma_2).
\end{equation}

Wave functions for the $N$ fermions are sections of the lift $\widehat E$ to
$\covspa$ of the bundle $E$ over $\Q$ with fiber
\begin{equation}\label{nw}
E_q=\bigotimes_{\canq \in q} W
\end{equation}
and  (nontrivial) connection inherited from the trivial connection
on $M\times W$. If the dynamics for $N=1$ involves wave functions on
$\widehat{M}$ obeying (\ref {e:percon06}) with topological factor
$\Gamma_\deckt(\hat{\boldsymbol{q}})=\Gamma_\deckt$ given by a
unitary representation of $\fund{M}$ (i.e., independent of
$\hat{\boldsymbol{q}}$), then the $N$ fermion wave function
obeys (\ref{e:percon06}) with topological factor

\begin{equation}\label{biggamma}
\Gamma_{\sigma}(\covq)=\mathrm{sgn}(p)\bigotimes_{\canq \in
\pi(\covq)}\Gamma_{\sigma^{(i_{\covq}(\canq))}}
\equiv \mathrm{sgn}(p)\Gamma_{\tilde\sigma}(\covq)
\end{equation}
where for  $\covq=(\hat\canq_1,\dots,\hat\canq_N), \
\pi(\covq)=\{\pi_M(\hat\canq_1),\dots,\pi_M(\hat\canq_N)\}$ and
$i_{\covq}(\pi_M(\hat\canq_j))=j$. Since

\begin{equation}\label{prod2}
\Gamma_{\tilde\sigma_1 \tilde\sigma_2}(\covq) = \Gamma_{\tilde\sigma_1}
(\covq) \, \Gamma_{\tilde\sigma_2} (\covq)
\end{equation}
we find, using (\ref{sprod}) and  (\ref{prod2}), that

\begin{subequations}\label{htr}
\begin{align}
\Gamma_{\sigma_1 \sigma_2}(\covq)&=
\mathrm{sgn}(p_1p_2)\Gamma_{p_2^{-1}\tilde\sigma_1
p_2\tilde\sigma_2}(\covq)\\ &=\mathrm{sgn}(p_1)\Gamma_{p_2^{-1}\tilde\sigma_1
p_2}(\covq)\mathrm{sgn}(p_2)\Gamma_{\tilde\sigma_2}
(\covq)\\ &=P_2\Gamma_{\sigma_1}(\covq)P_2^{-1}\Gamma_{\sigma_2}
(\covq),
\end{align}
\end{subequations}
which agrees with (\ref{twistedrep}) since the holonomy on the bundle $E$
is given by permutations $P$ acting on the tensor product (\ref{nw}).

\section{The \asp}
\label{sec:asp}

\z{We have seen that for a Riemannian manifold $\Q$ that is multiply
connected, there are additional possibilities for a Bohmian dynamics beyond
the usual ones. These new possibilities correspond to (twisted)
representations of $\fund{\Q}$, the most important of which are given by
the characters. In fact, unless the potential $V$ is very special, the
characters are the only representations that define a possible dynamics
involving that potential.} 

\z{We summarize our discussion so far, invoking the special status of the
characters, in the}

\noindent \textbf{\asp}.  \textit{Consider a quantum system whose
configuration space is given by the Riemannian manifold $\gencon$ and whose
value space for the wave function is given by the the Hermitian vector space
$\sspa$ [or the \herm\ bundle $E$ over $\Q$]. Then for every character
$\gamma$ of the fundamental group $\fund{\gencon}$, there is a family
${\cal B}_\gamma=\{{\cal B}_\gamma(V)\}$ of Bohmian dynamics, one for each
potential, i.e., Hermitian function $V: \gencon \to \sspa \otimes \sspa^*$
[or Hermitian section $V$ of $E \otimes E^*$]. The dynamics ${\cal
B}_\gamma(V)$ associated with the potential $V$ can be taken to be defined
by
\begin{subequations} \label{e:asp} \begin{align} \psi (\deckt \covq) &=
\concov \psi(\covq)\,,\\ i\hbar \frac{\partial \psi}{\partial t} (\covq) &=
-\tfrac {\hbar^2} {2} \gD \psi(\covq) + \wh{V}(\covq) \, \psi(\covq)\,,\\
\frac{dQ_t}{dt} &= \hbar \, \proj^* \Bigl( \im \, \frac{(\psi,\nabla
\psi)}{(\psi,\psi)} \Bigr) (Q_t) \end{align} \end{subequations} with $\psi
\in L^2(\covspa,\sspa,\gamma)$ [or $\psi \in L^2(\covspa, \wh{E},
\gamma)$].}

\medskip

Equations \eqref{e:asp} are identical with \eqref{e:percon},
\eqref{e:sch06}, and \eqref{e:bohm06}. Recall that the characters of
$\fund{\Q}$ are canonically identified with those of $\deckg$.

\z{The \asp\ corresponds to the symmetrization postulate for the case of $N$
identical particles in $\RRR^3$: in this case the natural configuration
space $\Q=\nrtre$ and the fundamental group $\fund{\Q}=S_N$, the group of
permutations of $N$ elements, which has two characters, the trivial
character, corresponding to bosons, and the alternating character,
corresponding to fermions.}


We now wish to elaborate upon why the theories given by characters deserve
special attention, and are, arguably, the only possibilities for theories
that can be regarded as fundamental. There are at least four crucial
considerations:  (i)~freedom, (ii)~genericity,  (iii)~theoretical
stability, and (iv) irreducibility.

 It seems within our power to expose a physical system, for example a
system of $N$ identical particles (which has the multiply-connected
configuration space $\nrtre$), to a wide variety of potentials. As we have
noted already in Remarks \eqref{rmk:ch}, \eqref{rmk:gen}, \eqref{rmk:gen2},
[and \eqref{rmk:chb}], if we can arrange any potential we like, or if the
potentials we can arrange are sufficient to generate together the algebra
$\mathrm{End}(\sspa)$ [respectively $\mathrm{End}(E_q)$], then the (twisted)
representations defining the dynamics must be given by a character.  For
example, as we show in \cite{topid2}, if we can expose a system of $N$
identical particles to arbitrary magnetic fields $\magnetic$ then the
potentials \eqref{pauli2} on the natural spin bundle \eqref{spinbundle}
over $\nrtre$, which occur in the Pauli equation, generate
$\mathrm{End}(E_q)$.

The second consideration is based on the hypothesis that, to the extent that
a Hamiltonian defining a fundamental physical theory can be regarded as a
Schr\"odinger operator $\rawH$, the potential $V$ is rather
generic, or at least not too special. But generically we have that
$\mathrm{Alg}(V(\Q)) = \mathrm{End}(\sspa)$ [and that $\mathrm{Alg}(V(\Q)_q
\cup \Theta_q) = \mathrm{End} (E_q)$].  It then follows, as
we have pointed out in Remarks \ref{rmk:gen} and \ref{rmk:genb}, that the
dynamics belongs to $\class_1$. And even systems that we can describe to a very
good degree of approximation by special (e.g., scalar) potentials [and, if
appropriate, special (e.g., flat) \herm\ bundles] then still cannot have a
dynamics from $\class_2 \setminus \class_1$.

The third consideration concerns the stability of the theory under
perturbations. The idea is that the theoretical description of a system
(such as, again, $N$ identical particles) should not be so delicately
contrived as to make sense only for a single potential $V$, but should also
make sense for all potentials close to $V$. (One reason why one might
require theoretical stability is the idea that our theoretical descriptions
may be idealized, for example when we take physical space to be Euclidean
$\RRR^3$ or a magnetic field to vanish, neglecting small perturbations.)
This implies that the theory should be well defined for a generic
potential, allowing only dynamics of class $\class_1$.

Finally, and perhaps most importantly, it seems reasonable to demand of a
fundamental physical theory that it be suitably irreducible. But it follows
from Remark \eqref{rmk:gen} [respectively \eqref{rmk:genb}] that a (twisted)
representation can fail to be given by a character only when
$\mathrm{Alg}(V(\Q)) \neq \mathrm{End}(\sspa)$ [respectively when
$\mathrm{Alg}(V(\Q)_q \cup \Theta_q) \neq \mathrm{End} (E_q)$], and in this
case the Schr\"odinger dynamics is decomposable into a direct sum of
dynamics corresponding to subspaces of the value-space $\sspa$ [or to
sub-bundles].  One might wonder, in this case, why the full value-space [or
bundle] was involved to begin with.

These considerations are of course related. For example, a generic
potential clearly corresponds to an irreducible dynamics. Freedom
relies on genericity in the following way. Since our one universe has in
fact just one Hamiltonian and thus just one potential $V=V_\mathrm{univ}$,
what must be meant when one speaks of exposing a system to various
potentials $V_\mathrm{sys}$ is that
\begin{equation}
  V_\mathrm{sys} (q_\mathrm{sys})  = V_\mathrm{univ}
  (q_\mathrm{sys}, Q_\mathrm{env})\,,
\end{equation}
for all configurations $q_\mathrm{sys}$ of the system, where
$Q_\mathrm{env}$ is the actual configuration of the environment of the
system (i.e., the rest of the universe), which we can control to a
certain extent. In words, $V_\mathrm{sys}$ is the \emph{conditional
potential} of a subsystem of the universe. Thus, the diversity of
potentials that we can arrange for a system is inherited from the
richness of the potential of the universe: if $V_\mathrm{univ}$ were
scalar, we would be unable to arrange potentials $V_\mathrm{sys}$
other than scalars. Thus, the origin of the freedom of potentials must
lie in genericity. On the other hand, freedom, since it requires the
genericity hypothesis, lends support to it.

\section{Conclusions}
\label{sec:conclusions}

We have studied the possible quantum theories on a topologically
nontrivial configuration space $\Q$ from the point of view of Bohmian
mechanics, which is fundamentally concerned with the motion of matter
in physical space, represented by the evolution of a point in
configuration space.

Our goal was to find, define, and classify all Bohmian dynamics in \gencon,
where the wave functions may be sections of a \herm\ vector bundle
$E$. What ``all'' Bohmian dynamics means is not obvious; we have followed
one approach to what it can mean; other approaches are described in
\cite{topid1B, topid1C,topid1D}. The present approach uses \z{ wave functions
$\psi$} that are defined on the universal covering space $\covspa$ of
\gencon\ and satisfy a periodicity condition ensuring that the Bohmian
velocity vector field on $\covspa$ defined in terms of $\psi$ can be
projected to \gencon. We have arrived in this way at two natural classes
$\class_1 \subseteq \class_2$ of Bohmian dynamics beyond the immediate
Bohmian dynamics.  A dynamics from $\class_1$ is defined by a potential and
some topological information encoded in a character (one-dimensional
unitary representation) of the fundamental group of \z{the} configuration space,
$\fund{\Q}$.  A dynamics from $\class_2$ is defined by a potential and a
more general algebraic-geometrical object, a ``periodicity section''
$\Gamma$.

The dynamics of $\class_2 \setminus \class_1$ exist only for special
potentials. Those of $\class_1$, however, are compatible with \emph{every}
potential, as one would desire for \z{what} could be \z{considered} a
version of quantum mechanics in $\Q$.  We have thus arrived at the known
fact that for every character of $\fund{\Q}$ there is a version of quantum
mechanics in $\Q$; we have formulated this in terms of Bohmian mechanics as
the ``\asp.'' A consequence, which will be discussed in detail in a sister
paper \cite{topid2}, is the symmetrization postulate for identical
particles. These different quantum theories emerge naturally when one
contemplates the possibilities for defining a Bohmian dynamics in $\Q$.

\section*{Acknowledgments}

We thank Kai-Uwe Bux (Cornell University), Frank Loose
(Eberhard-Karls-Universit\"at T\"ubingen, Germany) and Penny Smith
(Lehigh University) for helpful discussions.

R.T.\ gratefully acknowledges support by the German National Science
Foundation (DFG) through its Priority Program ``Interacting Stochastic
Systems of High Complexity'', by INFN, and by the European Commission
through its 6th Framework Programme ``Structuring the European Research
Area'' and the contract Nr. RITA-CT-2004-505493 for the provision of
Transnational Access implemented as Specific Support Action.  N.Z.\
gratefully acknowledges support by INFN. The work of S.~Goldstein was
supported in part by NSF Grant DMS-0504504.

Finally, we appreciate the hospitality that some of us have enjoyed,
on more than one occasion, at the Mathematisches Institut of
Ludwig-Maximilians-Universit\"at M\"unchen (Germany), the Dipartimento
di Fisica of Universit\`a di Genova (Italy), the Institut des Hautes
\'Etudes Scientifiques in Bures-sur-Yvette (France), and the
Mathematics Department of Rutgers University (USA).

\end{document}